\documentclass[review]{elsarticle}

\usepackage[a4paper, margin=2.5cm]{geometry}

\usepackage{lineno,hyperref}
\usepackage{numprint}
\npthousandsep{,}\npthousandthpartsep{}\npdecimalsign{.}
\modulolinenumbers[5]

\journal{Fire Safety Journal}

\usepackage{color}
\usepackage{comment}
\usepackage{enumitem}
\usepackage{tabularx}
\usepackage{array}
\usepackage{color, colortbl}
\usepackage{multirow}
\usepackage{longtable}
\usepackage{pdflscape}
\usepackage{amssymb}
\usepackage{subfigure}
\usepackage{amsmath}
\usepackage{lineno}
\usepackage{amsmath}
\usepackage{amssymb}
\usepackage{amsfonts}
\usepackage{algorithmic}
\usepackage{algorithm}
\usepackage{hyphenat}

\definecolor{Gray}{gray}{0.9}

\definecolor{mygreen}{RGB}{0, 150, 51}

\newcommand{\quotes}[1]{``{#1}''}

\usepackage{fancyhdr}
\begin{document}

\newcolumntype{R}[1]{>{\raggedright\let\newline\\\arraybackslash\hspace{0pt}}m{#1}}
\newcolumntype{C}[1]{>{\centering\let\newline\\\arraybackslash\hspace{0pt}}m{#1}}
\newcolumntype{L}[1]{>{\raggedleft\let\newline\\\arraybackslash\hspace{0pt}}m{#1}}

\begin{frontmatter}
\title{Experimental data about the evacuation of preschool children from nursery schools -- Part II: Movement characteristics and behaviour\footnote{This is a preprint version of the article accepted to Fire Safety Journal with DOI 10.1016/j.firesaf.2023.103797}}

\pagestyle{fancy}

\author[fsv,uceeb]{Hana Najmanov\'a}
\ead{hana.najmanova@cvut.cz}
\author[lund]{Enrico Ronchi}
\ead{enrico.ronchi@brand.lth.se}
\cortext[cor1]{Corresponding author. Tel. +420224357151, email: hana.najmanova@cvut.cz}
\address[fsv]{Czech Technical University in Prague, Faculty of Civil Engineering, Thakurova 7, 166 29 Prague, Czechia}
\address[uceeb]{Czech Technical University in Prague, University Centre for Energy Efficient Buildings, T\v rineck\'a 1024, 273 43 Bu\v st\v ehrad, Czechia}
\address[lund]{Lund University, Department of Fire Safety Engineering, John Ericssons v\"ag 1, 22363, Lund, Sweden}

\begin{abstract}
This article presents experimental data sets about the evacuation movement and behaviour of preschool children observed during 15~evacuation drills in 10~nursery schools in the Czech Republic involving 970~children (3–-7~years of age) and 87~staff members. The movement characteristics and behaviour of the children were studied in 100~measurement areas defined on evacuation routes in corridors, on staircases, and in doorways. A total of \numprint{23000}~data points were gathered to investigate travel speeds, specific flows, densities, and their relationships, as well as movement behaviour and procedures during the evacuation drills. Variables of interest included the levels of physical assistance provided to children, hand holding behaviour, use of handrails, and footstep patterns on staircases (i.e., \quotes{marking time} patterns). The observations revealed that the movement characteristics of preschool-age children were age-dependent. On horizontal routes, higher travel speeds were measured in corridors (mean value 1.9~m$\cdot$s$^{-1}$) than for landings of straight staircases (mean value 0.96~m$\cdot$s$^{-1}$) and doorways (mean value 0.99~m$\cdot$s$^{-1}$). Nine behavioural statements summarising the evacuation movement behaviour of preschool children observed in this study are presented in order to enrich future applicability of the findings in engineering applications (e.g., evacuation modelling).

\end{abstract}
\begin{keyword}
Fire safety \sep
Evacuation \sep
Preschool children \sep
Movement \sep 
Travel speed \sep 
Experimental drill 
\end{keyword}

\end{frontmatter}
\thispagestyle{fancy}

\paragraph{Highlights} \par

\begin{itemize}
	\item Experimental data on evacuation movement of children 3--7~years of age are presented. 
    \item 15~evacuation drills in ten nursery schools with 970~children (\numprint{23000} data points).
    \item Data includes travel speeds, flows, and densities in corridors, staircases, doorways.
    \item The movement behaviour of children is presented as behavioural statements.
\end{itemize}

\newpage
\section{Introduction}\label{sec:intro}
The evaluation and prediction of movement and behaviour of building occupants represent key elements in the fire safety and evacuation design process. Taking into account the wide range of capabilities of human movement, differences in individual movement characteristics must be carefully assessed with particular importance placed on the characteristics of vulnerable populations with limited levels of self-rescue capabilities \cite{bukvic_review_2021}. Young children can be considered to be \quotes{vulnerable} due to their developmental processes, which can result in limited motor and cognitive skills \cite{payne_human_2012}. During early childhood (from approximately two to six years of age), fundamental movement skills development progresses rapidly \cite{adolph_development_2017}. Children typically begin to walk without support at around one year; however, changes in their gait patterns attributed to growth can last up to the age of four, with matured gait patterns emerging at seven \cite{hillman_development_2009}. Walking speed and the length of steps children take are affected most by physical size; that is, the length of their legs \cite{ malina_motor_2004}. Running attempts occur between two to three years of age, due to the development of dynamic balance, core strength, and multilimb coordination, all of which improved until the age of six or seven \cite{gallahue_understanding_2010}. For preschool-age children, ascending and descending stairs is biomechanically challenging due to increased demands on foot placement and dynamic balance skills \cite{silverman_whole-body_2014}. Marking time (i.e., placing both feet on the same step) is usually replaced by feet alternation from three to four years of age; during this time, children may need handrails or other support when walking downstairs \cite{williams_stair_1994}. In engineering design, \quotes{childhood} should not be therefore perceived as a homogeneous term and any evaluation of children's movement abilities and behaviour should be linked to a specific level of child development in the simplest form related to age \cite{singer_piaget_1996}. \par
To explore the specifics of the evacuation movement of preschool-age children, several research studies on the evacuation movement of children under six years of age have been performed in the field of fire safety and pedestrian and evacuation dynamics \cite{murozaki_study_1985,kholshevnikov_pre-school_2009,kholshevnikov_study_2012,capote_children_2012,cuesta_exploring_2013,takizawa_study_2013,taciuc_determining_2014,larusdottir_evacuation_2014,cuesta_collection_2016,najmanova_experimental_2017,hamilton_toward_2019,fang_experimental_2019,li_comparative_2020, yao_research_2020, yao_childrens_2021}. 
However, as recent research studies (such as \cite{larusdottir_evacuation_2014,najmanova_experimental_2017, hamilton_toward_2019, li_comparative_2020}) have indicated, understanding the evacuation dynamics of preschool children is a complex issue, and more scientific work is needed to support and extend our knowledge and to enrich the engineering databases used in fire safety design.  %

This article presents part of an experimental study investigating the evacuation of children three to seven years of age from early childhood educational centers (hereafter referred to as \quotes{nursery schools}). This study delves into new experimental data sets and information regarding 15~evacuation drills involving 970~children and 87~staff members in ten participating nursery schools in the Czech Republic (May and June 2019). Due to the large scope of research and the substantial amount of data collected, the findings of this study are presented in two separate publications that follow the engineering timeline for the evacuation of buildings from fires \cite{purser_behaviour_2003,ng_brief_2006}: the pre-movement phase (Part I, first paper) and the movement phase (Part II, this paper) of the evacuation process. 
This paper reports on the results obtained regarding the movement characteristics and behaviour of participating children during the movement phase of the evacuation drills.
A deeper understanding of the specifics of children's movement patterns and behaviour is provided by detailed experimental data sets interpreted, taking the full context of the research methods used into consideration in order to define a set of nine behavioural statements \cite{gwynne_guidance_2016,kuligowski_guidance_2017}. To enable appropriate engineering applications, the elements of physical movement characteristics (including the concept of occupied areas of children) and behavioural itineraries are also investigated.

\section{Research methods}
\label{sec:methods}
The paper presents experimental data sets collected in 15 evacuation drills conducted in ten participating nursery schools. To cover the diversity of evacuation conditions and to contribute to the overall knowledge on preschool children evacuation, the study includes a large variance in the participating subjects and conditions (see Section 2.1). This means that further data collection in comparable settings/conditions are likely to be in a similar range what has been presented.
This section provides the experimental background of the study including a basic description of the participants, the variables observed, and the data collection and analysis methods used (for more details please refer to \cite{najmanova_evacuation_2020}). \par 

\subsection{Participants and ethical considerations}\label{subsec:parti}
A basic description of the ten participating nursery schools including their building features, background information about the evacuation drills and occupancy is summarised in Table~\ref{tab:schools}. In addition to different building designs and capacities, the participating nursery schools also varied in their experience with drills and evacuation procedures. As a result, the boundary conditions of the evacuation drills differed considerably. 
\begin{table}[!hbt]
    \footnotesize
    \centering
    \begin{tabular}{R{1cm}|R{1.0cm}|R{2.2cm}|R{1.5cm}|R{2.2cm}|R{1.1cm}|R{1.1cm}|R{1.4cm}|R{0.8cm}}
      \hline
   
      \multirow{2}{=}{\textbf{School} \newline \textbf{label}}  & \multicolumn{2}{l|}{\textbf{Building description}} & \multicolumn{2}{l|}{\textbf{Evacuation drills}} & \multicolumn{3}{l}{\textbf{Occupancy}}  \\
       \cline{2-9}
    
       &  \textbf{Floors} & \textbf{Used staircase} & \textbf{Serial No. of drill} & \textbf{Evacuation type\textsuperscript{1)}}  & \textbf{Class type} & \textbf{Classes} & \textbf{Children} & \textbf{Staff}\\
       \hline
     A &   3 & External spiral & 1 & Announced & HET & 4 & 74 & 5 \\
     \hline
     B &  2 & Internal straight & 1 & Unannounced & HOM, HET & 11 & 207 & 21 \\
     \hline
     C  & 2 & Internal straight & 1 & Semi-announced &  HOM & 4 & 75 & 5 \\
     \hline
     \multirow{2}{=}{D} & \multirow{2}{=}{1} & \multirow{2}{=}{--} & 1 & Announced & HET & 4 & 63 & 4 \\
     \cline{4-9}
      &  &  & 2 & Semi-announced & HET & 4 & 61 & 4 \\
     \hline
     E &  2 & Internal straight & 1 & Unannounced & HOM & 6 & 100 & 9 \\
     \hline
     \multirow{2}{=}{F} & \multirow{2}{=}{2} & \multirow{2}{=}{Internal straight} & 1 & Announced & HET & 1 & 20 & 2 \\
     \cline{4-9}
      &  & & 2 & Announced & HET & 1 & 23 & 2 \\
    \hline
     \multirow{2}{=}{G} & \multirow{2}{=}{1} & \multirow{2}{=}{--} & 1 & Announced & HET & 3 & 52 & 4 \\
     \cline{4-9}
      &  & & 2 & Semi-announced  & HET & 3 & 50 & 3 \\
    \hline
     \multirow{2}{=}{H} & \multirow{2}{=}{2} & \multirow{2}{=}{Internal and external straight} & 1 & Announced & HOM & 4 & 61 & 4 \\
     \cline{4-9}
      &  & & 2 & Unannounced & HOM & 2 & 24 & 3 \\
    \hline
         \multirow{2}{=}{I} & \multirow{2}{=}{1} & \multirow{2}{=}{--} & 1 & Announced & HET & 1 & 12 & 3 \\
     \cline{4-9}
             & &  & 2 & Semi-announced & HET & 1 & 12 & 3 \\
    \hline
    J &  3 & Internal straight & 1 & Announced & HOM & 8 & 136 & 16 \\
    \hline
    \multicolumn{6}{l}{\textbf{Total}} & \textbf{57} & \textbf{970} & \textbf{87}  \\
    \hline
     \multicolumn{9}{l}{\footnotesize \textsuperscript{1)} See Section 2.2 for details}  \\
    \multicolumn{9}{l}{\footnotesize{HOM: homogeneous class, HET: heterogeneous class} } \\
    \hline
    \end{tabular}
    \caption{Overview of the participating nursery schools.}
    \label{tab:schools}
\end{table}

The participating children attended both homogeneous classes (those with similarly-aged children) and heterogeneous classes (where children's ages ranged from 3~to 6~years). For the purposes of this study, four different age groups of children were defined: Junior (3--4~years), Senior (5--6~years), Senior+ (6--7~years), and Mixed children (3--6~years); see Table~\ref{tab:age}. Due to ethical considerations, the researchers only collected the age ranges related to entire classes.  
\begin{table}[hbt!]
    \centering
    \footnotesize
    \begin{tabular}{R{3cm}|R{3cm}|R{3cm}|R{3cm}}
        \hline
   
       \textbf{Age group [years]} & \textbf{Label in this study} & \textbf{Number of classes} & \textbf{Number of children} \\
        \hline
        3--4 & Junior & 17 & 265 \\
        \hline
        5--6 & Senior & 15 & 295 \\
        \hline
        6--7 & Senior+ & 1 & 22 \\
        \hline
        3--6 & Mixed & 24 & 389 \\
        \hline
       \multicolumn{1}{l}{} & \multicolumn{1}{l}{\textbf{Total}} & \multicolumn{1}{l}{\textbf{57}} & \multicolumn{1}{l}{\textbf{970}} \\
       \hline
    \end{tabular}
    \caption{Age-related overview of the participating children.}
    \label{tab:age}
\end{table}
Taking into account fundamental ethical principles in research involving human subjects, particular care was taken to ensure appropriate ethical considerations and to guarantee the participants were protected and respected \cite{nilsson_exit_2009}. The research project was approved by the Czech Technical University in Prague’s ethical review board. In the planning phase of the study, the nursery school directors were invited to sign an informed consent and communication with the parents of the children was established.  Due to the large number of children involved in the study, parental permission was obtained as opt-out consent, that is, passive approval. Parents received information on the coming evacuation drills and video recordings in advance and could withdraw participation of their child in the study by contacting the school director or the authorised nursery school staff member. Staff members signed informed consent forms to use the data before or immediately after an experiment (the latter, if an evacuation drill was not announced in advance).
\subsection{Experimental setup and limitations}\label{subsec:setup}
To ensure the well-being and safety of the participants and to maintain an acceptable level of (semi-) naturalistic conditions, experimental control, and repeatability, the observation of the evacuation drills were based upon the evacuation procedures of each respective nursery school. The study was entirely observational and the researchers did not intervene in the nursery schools' evacuation procedures. Nursery school directors decided on the amount of information about evacuation drills provided to staff and children before the experiments (also referred to as \quotes{Level of announcement} in this study). In the announced evacuation drills, the date and time of the evacuation drills were made available to all staff members prior to the drills. In semi-announced evacuation drills, evacuation drills were also announced to all staff members beforehand; however, some part of the information was not provided to them (e.g., date or/and time of the evacuation drill). No information on evacuation drills was provided to staff members prior to unannounced evacuation drills. The amount of information provided to children by parents and staff members was not controlled; only in the case of unannounced drills, parents were asked not to communicate about the planned drill with the children. Evacuation drills were recorded using outdoor digital cameras (resolution $720\times480$, frame rate 30~fps) temporarily installed on evacuation routes using different mounting equipment (e.g., suction cups, flexible tripods). To maintain the unannounced nature of the experiments, cameras were installed only in the corridors for unannounced evacuation drills. In addition, preparation works were scheduled to minimise the contact between researchers and uninformed staff members. If necessary, the school director decided on an explanation for excusing the unusual activities in the building. The video footage was synchronised, cut, and stored within an encrypted device. Data were processed using frame-by-frame analysis in Avidemux 2.7.3 software and organised into anonymous spreadsheets. \par
Regarding limitations of this study, since the semi-naturalistic conditions of the evacuation drills could not inherently capture the diversity of real evacuation events, their potential impacts on evacuation movement and behaviour of participants presented cannot be generalised to other contexts. The data sets presented are affected by uncertainties related to experimental measurements and statistical assumptions. The analysis method used for flow calculations generated a set of discrete jumps in the data points, which can be mainly visible in several flow-density scatter plots. The data sets collected on certain parts of the escape routes (e.g., spiral stairs) and in certain age groups (e.g., only one class of Senior+ children) may also not be fully representative due to the limited number of observations. Furthermore, movement characteristics and behaviour were observed mainly under low density conditions in the evacuation drills; therefore, the reported data sets lack experimental observations in high-density scenarios.
\subsection{Observed variables}
The evacuation behaviour of children and staff members was observed during their movement through school buildings and observed variables included the evacuation procedures used, levels of physical assistance provided to children, and children's hands holding. Special attention was paid to children's movement on staircases where the use of handrails (and , in some cases, searching other type of support) and feet alternation (marking time) patterns were monitored. Considering movement characteristics of children, three fundamental variables that describe human movement and their relationships were analysed: travel speed, specific flow, and density. \par
Travel speeds were measured during continuous movement of children (i.e.,~not influenced by the acceleration or deceleration phases, waiting, or queuing) and were therefore called \quotes{movement speeds}. In several cases, people stopped inside measurement areas, and their movement speeds were calculated excluding the waiting times in the measurement area. Children's walking and running speeds were distinguished on horizontal parts of evacuation routes (corridors, landings of straight staircases, and doorways) based on the presence of a flight phase (i.e., both feet in the air) observed by analysis of video recordings. Specific flows were calculated at the exits of predetermined measurement areas at each second interval when these components were used, i.e., knowing the times when study participants crossed observation checkpoints. In corridors and on staircases, the width of the observed checkpoint was assumed to be the width of a measurement area. For movement through doors the width of an observation checkpoint was the clear width of the door opening, i.e.,~without any boundary layers. Specific flows were analysed without distinction for children who were walking or running, since both kinds of movement could occur simultaneously in a measurement area. \par  
The density of children $D$ [pers$\cdot$m$^{-2}$ or m$^{2}$\,m$^{-2}$] was calculated as the ratio of the number of participants in the measurement area $N$ [pers or m$^{2}$] over the size of a measurement area [m$^2$]
\begin{equation}
    D=N \cdot A^{-1}
\end{equation}
The number of people in measurement area $N$ was calculated knowing the entrance and exit times for all those observed. Both staff members and children in the measurement areas were included in the number of people considered in density and flow calculations. The density assigned to each person was determined as an average value within the time period for when a participant moved within a measurement area. To provide comprehensive and comparable data sets, the density variable was expressed in both pers$\cdot$m$^{-2}$ as well as m$^{2}$\,m$^{-2}$ units \cite{predtechenskii_planning_1978} in this study \cite{najmanova_evacuation_2020}. This paper reports results using the units of m$^{2}$\,m$^{-2}$ calculated using the concept of occupied areas (horizontal projections of a child's body) based on the results of the 6\textsuperscript{th} Nationwide anthropological survey of children and adolescents in the Czech Republic \cite{kobzova_6th_2004}. The following values for areas occupied by children were considered: $A_{occ}=0.025~$m$^2$ for a Junior child (3--4~years of age), $A_{occ}=0.028~$m$^2$ for a Senior child (5--6~years of age), $A_{occ}=0.029~$m$^2$ for a Senior+ child, and $A_{occ}=0.026~$m$^2$ for a child in the Mixed classes (3--6~years of age). The occupied area by an adult was considered in a simplified way to equal as the occupied area for the observed children. This choice was made since only maximum one staff member at a time was observed in the measurement area. \par
\subsection{Measurement areas}\label{subsubsec:area}
\begin{figure}
    \centering
    \subfigure{\includegraphics[width=0.4\textwidth]{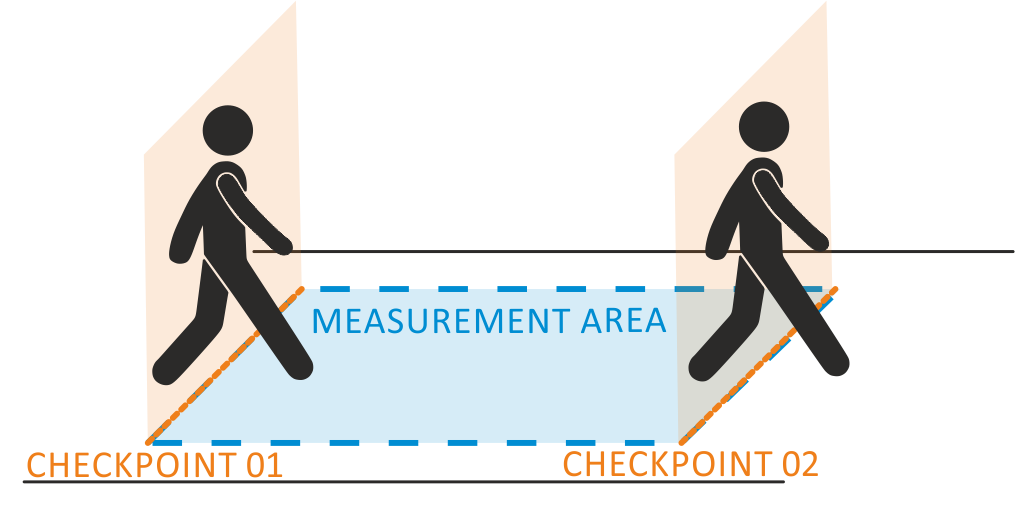}} 
    \subfigure{\includegraphics[width=0.4\textwidth]{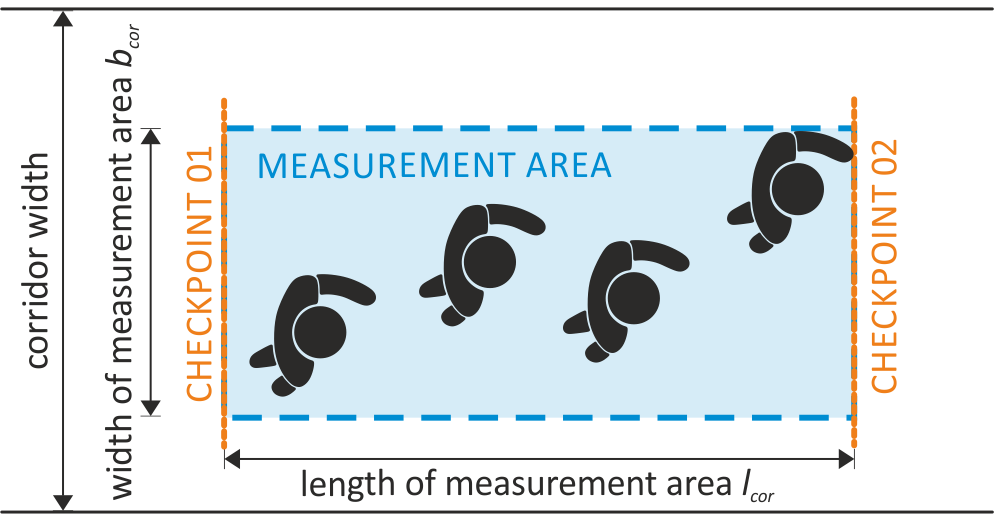}} 
    \caption{Measurement area in corridors.}
    \label{fig:cor}
\end{figure}

\begin{figure}[hbt!]
    \centering
    \subfigure{\includegraphics[width=0.44\textwidth]{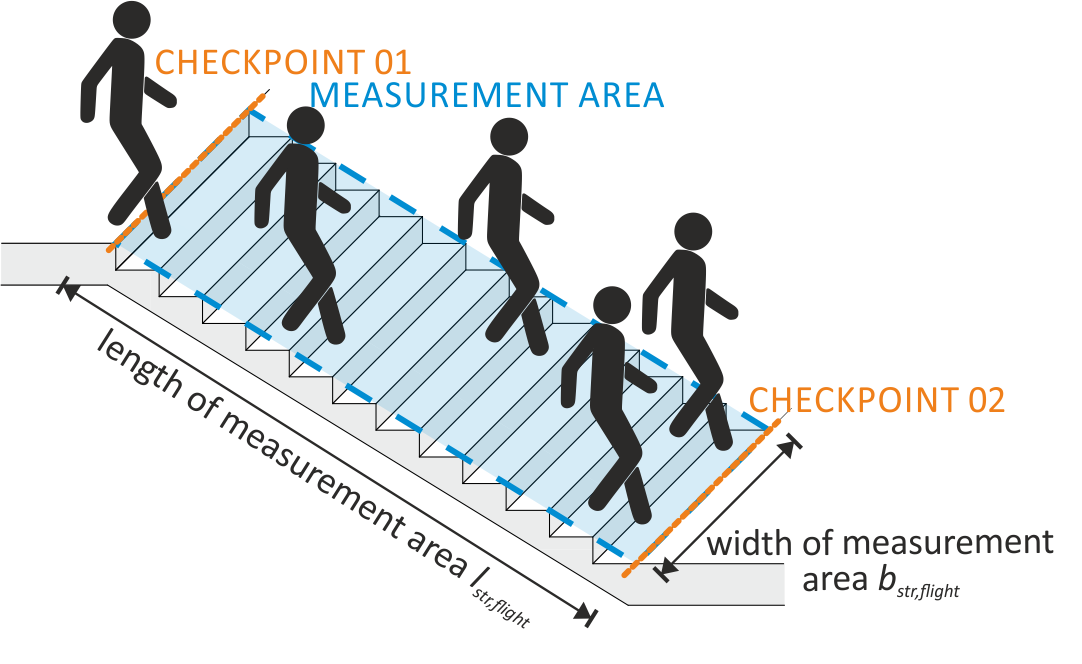}} 
    \subfigure{\includegraphics[width=0.44\textwidth]{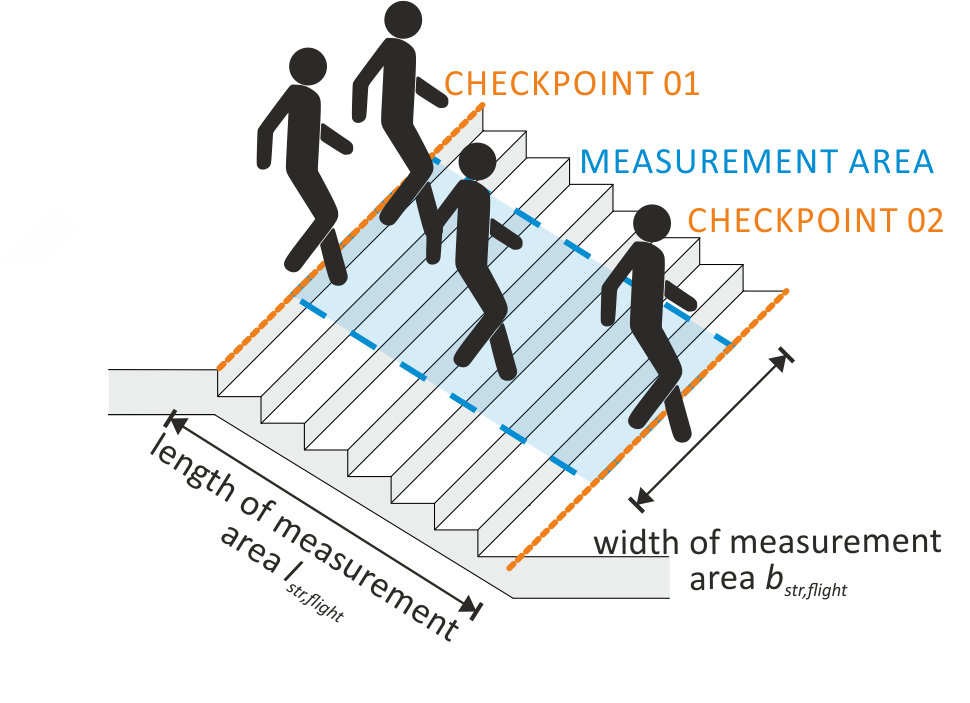}} 
    \caption{Measurement area on flights.}
    \label{fig:flight}
\end{figure}

\begin{figure}[hbt!]
    \centering
    \subfigure{\includegraphics[width=0.44\textwidth]{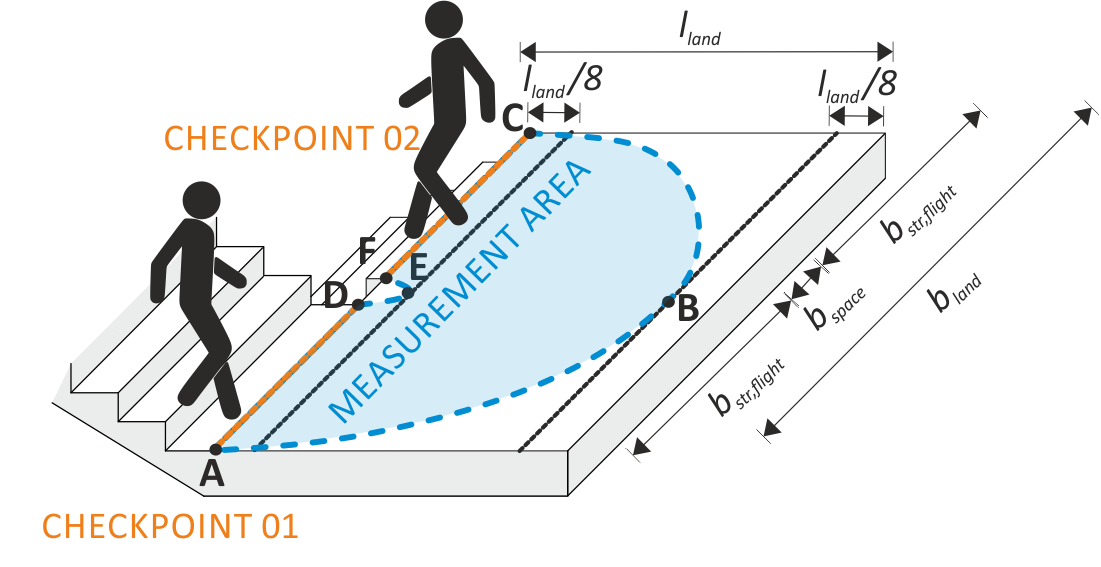}} 
    \subfigure{\includegraphics[width=0.44\textwidth]{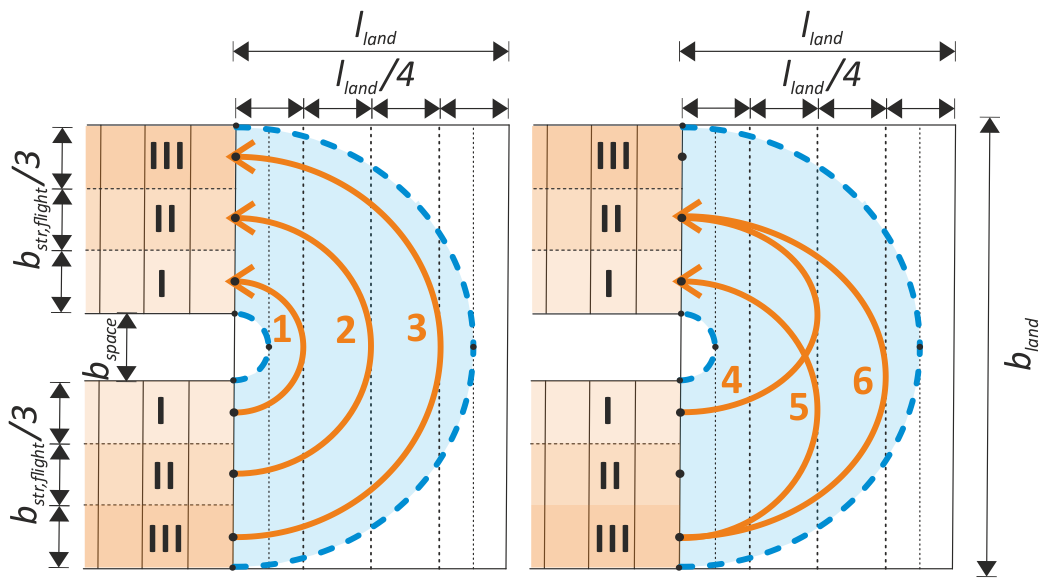}} 
    \caption{Measurement area on landings.}
    \label{fig:land}
\end{figure}

\begin{figure}[hbt!]
    \centering
    \subfigure{\includegraphics[width=0.42\textwidth]{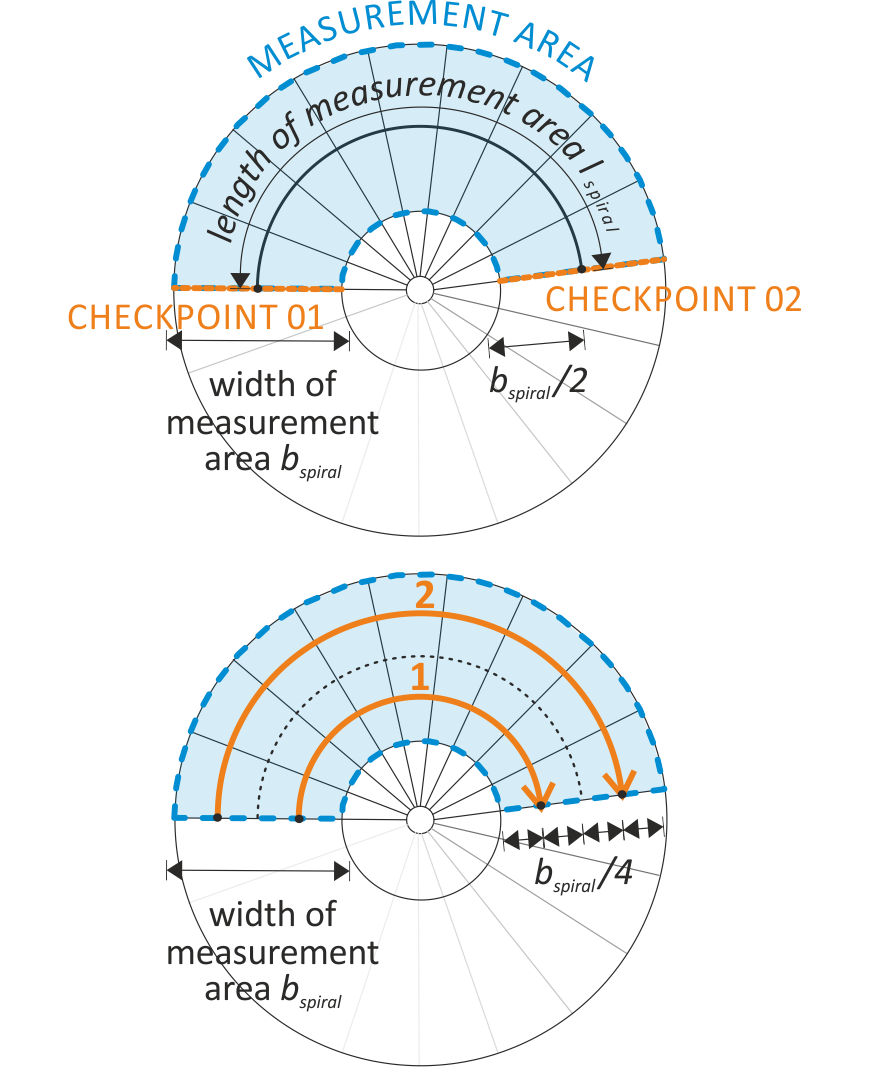}} 
    \subfigure{\includegraphics[width=0.42\textwidth]{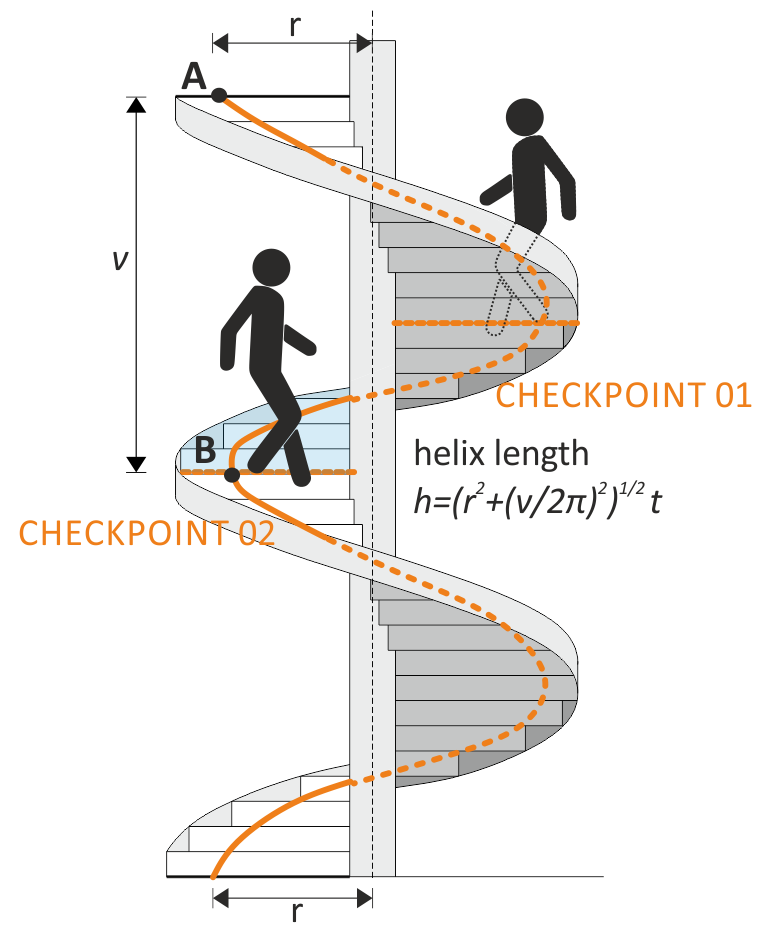}} 
    \caption{Measurement area on spiral staircases.}
    \label{fig:spiral}
\end{figure}

\begin{figure}[hbt!]
    \centering
    \subfigure{\includegraphics[width=0.35\textwidth]{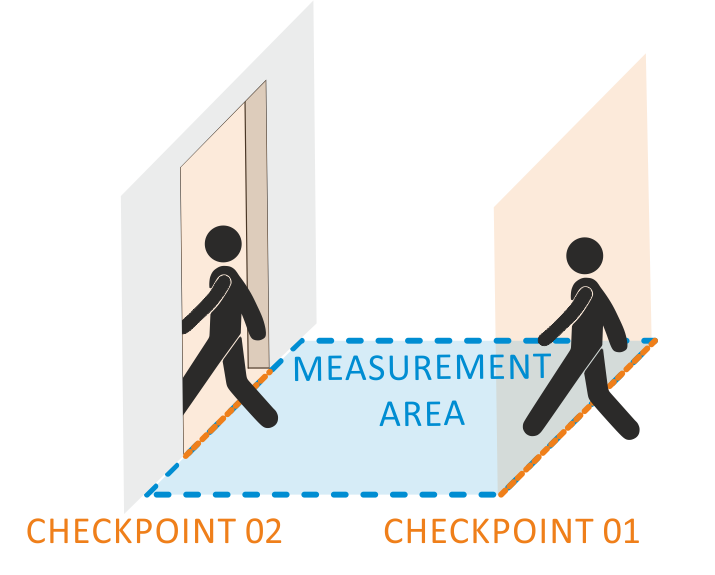}} 
    \subfigure{\includegraphics[width=0.35\textwidth]{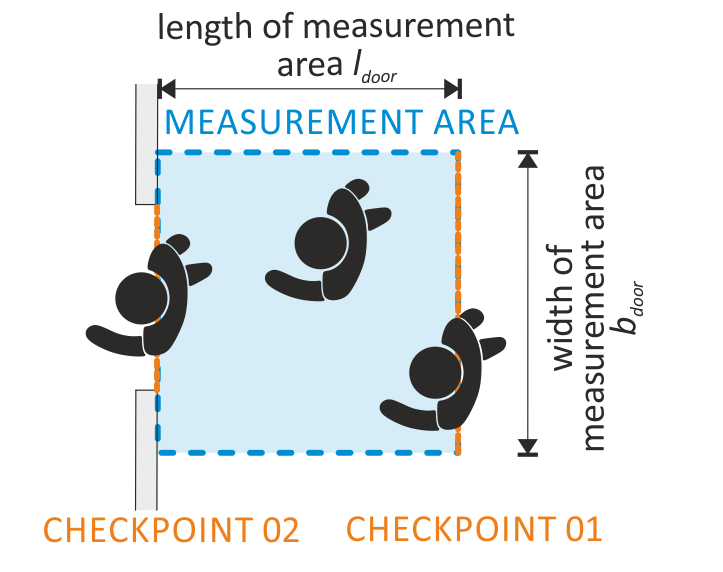}} 
    \caption{Measurement area in doorways.}
    \label{fig:door}
\end{figure}

To study the movement of participants through buildings, a total of 100~measurement areas were defined in corridors (see Figure~\ref{fig:cor}), on staircases (Figures~\ref{fig:flight},~\ref{fig:land} and \ref{fig:spiral}), and in doorways (see Figure~\ref{fig:door}). The entrances and exits for a measurement area were assigned as checkpoints, allowing us to collect the times when participants passed through measurement areas. In corridors and doors, participants were considered to have passed a checkpoint when the center of their bodies (simplified as shoulders) crossed the plane limited by the checkpoint. On staircases, it was assumed that participants entered or exited a measurement area when their first foot was flat on the first step in the measurement area (entering) or not on the last step in the measurement area (Figures~\ref{fig:cor}--\ref{fig:door}). \par 
The length of the measurement areas in corridors was 2~m, 3~m, or 5~m depending on the geometry of a corridor and a building's layout. The assumed width of a measurement area $b_{cor}$ [m] corresponded to the width of the evacuation route that the children used in the evacuation drills and was identified through the analysis of the video recordings. 
On straight staircases, movement was evaluated separately for flights and landings. The length of a measurement area for flights $l_{str,flight}$ [m] was defined as the length of a flight on the slope calculated as the number of steps multiplied by the distance between two steps on the slope. The width of a measurement area $b_{str,flight}$ [m] was the width that was used on the flight and was identified through the analysis of the video recordings. The measurement area on landings was defined as the area limited by two semi-ellipses (Figure~\ref{fig:land}). Aware of the considerable impact of the assumed travel distance on the calculated travel speeds, nine possible semi-ellipse shaped travel paths $d_{land}$ [m] (i.e., six different travel distances, see Arabic numbers in Figure~\ref{fig:land}) were assumed and assigned to participant individually using the analysis of video recordings. For this purpose \cite{ronchi_analysis_2014}, the assumed width of flights (i.e.,~entrance and exit parts landings) was divided into three subareas (see I, II, III in Figure~\ref{fig:land}) and the landing length was divided into four subareas. To provide results that neglect the influence of different parts of straight staircases and thus are suitable, e.g., for evacuation modelling, children's movement characteristics were also observed on entire staircases. Thus, a measurement area was assumed to be a complete staircase, including all flights and landings and travel distance was calculated individually for each participant as the sum of the travel distances on all flights and landings. Movement characteristics were calculated as weighted average values over time measured for particular flights and landings. \par

On spiral staircases, the length of a measurement area $l_{spiral}$ [m] was defined as the length of a section (approximately the half of one shift) of a circular helix calculated in the middle of the assumed width of the spiral staircase (Figure~\ref{fig:spiral}). The width of a measurement area $b_{spiral}$ [m] was assumed to be as the width used on the spiral staircase based on the analysis of the video recordings. 
Two possible travel paths were considered to more realistically evaluate participants' travel distances. \par 
\par

To evaluate movement through doorways, a measurement area was determined in front of a respective door (Figure~\ref{fig:door}). The area of each measurement area was fixed as 1~m$^2$, i.e.,~the length $l_{door}$ [m] and the width $b_{door}$ [m] of a measurement area were set to 1~m (Figure~\ref{fig:door}).
Further details of measurement areas, including dimensions and travel path calculations, can be found in \cite{najmanova_evacuation_2020}. \par

\section{Results} 
\label{sec:results}
This section presents the travel speed-density and specific flow-density relationships observed for the children during the evacuation drills. The data sets include fitting curves (trend lines with calculated coefficients of determination), confidence bounds of the mean response (95\% confidence level), and prediction bounds for a future observation (95\% confidence level) based on linear least squares regression analysis. To provide qualitative insight into preschool children’s evacuation dynamics, a simple linear regression analysis was used to determine the trend lines of observed data sets that followed the conventional shapes of curves known to be relevant in the literature (e.g.,~Society of Fire Protection Engineers, SFPE curves \cite{gwynne_employing_2016}): The trend lines for speed-density relationships were expressed using linear functions, and the flow-density data sets used polynomial functions. Although the trends indicated by the fitted statistical models used were signifficant at alpha level of 0.05 significant (p$<$0.0001), the calculated coefficients of determination are low and the provided trend lines have rather indicative purposes in many cases. This may be mainly attributed to a large dispersion of the observed data related to the variance of both pedestrian dynamics and evacuation conditions in the observed drills. In addition, tabulated results that reported mean, minimal, maximal, and standard deviation values for travel speeds and specific flows in different density intervals and for different age groups are summarised in \ref{app:cor}--\ref{app:door}). Raw data sets for the presented results can be found as supplementary material at \cite{najmanova_data_2023}. \par

\subsection{Movement in corridors} \label{subsec:cor}
\begin{figure}[hbt!]
    \centering
    \includegraphics[width=\textwidth]{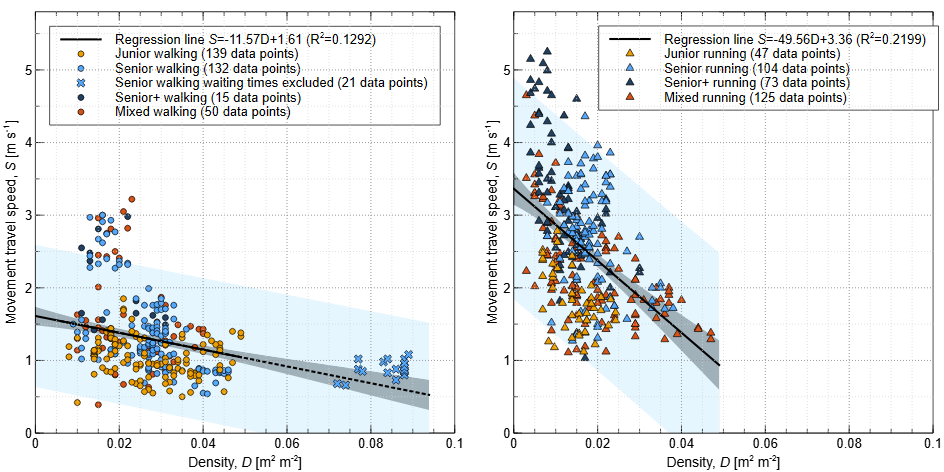} 
        \vspace{-8mm}
        \caption{
        \textbf{Left}: Walking travel speeds per density in corridors for different age groups;
        \textbf{Right}: Running travel speeds per density in corridors for different age groups.
    \textbf{Note}: Regression line, 95\% confidence bounds of the fitted curve (grey area), and 95\% prediction bounds for a future observation (blue area) are included. Dashed parts of the trend lines indicate density intervals for a limited number of data points.}
    \label{fig:speed_cor_walk+run}
\end{figure}

\begin{figure}[hbt!]
    \centering
    \includegraphics[width=\textwidth]{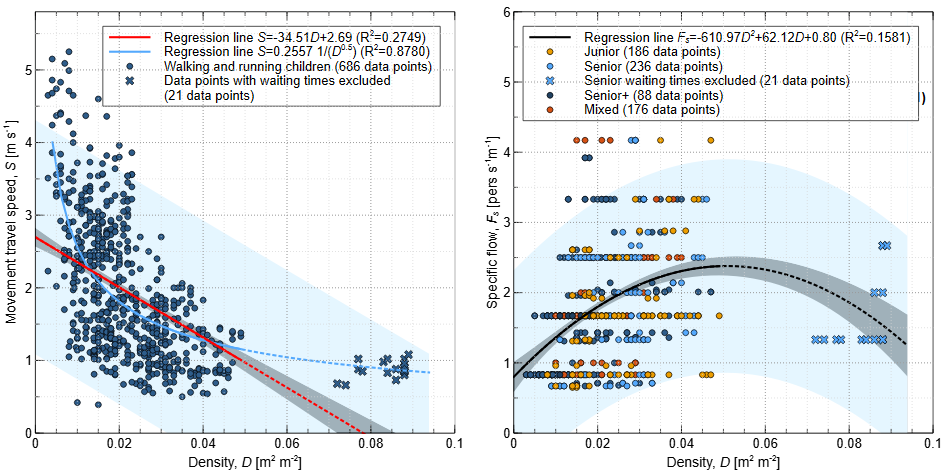} 
    \vspace{-8mm}
    \caption{
    \textbf{Left}: Travel speeds (walking and running combined) per density in corridors;  
    \textbf{Right}: Specific flows per density in corridors.
    \textbf{Note}: Regression line, 95\% confidence bounds of the fitted curve (grey area), and 95\% prediction bounds for a future observation (blue area) are included. Dashed parts of the trend lines indicate density intervals for a limited number of data points.}    \label{fig:speed_cor_total}
\end{figure}

Travel speeds, pedestrian flows, and densities in corridors were measured in 20~measurement areas in the seven participating nursery schools during eight experimental evacuation drills. %
Walking and running travel speeds were analyzed separately during the continuous movement of the children (i.e., movement travel speeds). One class (Senior class, 21~children, 21~data points) stopped in a measurement area, and thus their waiting times for this measurement area were excluded to calculate their travel speeds. 
Figure~\ref{fig:speed_cor_walk+run} shows the dependence of movement travel speed $S$ on density $D$ for children in different age groups (color coded) who were walking or running. The speed-density relationship for all observations (all age groups, walking and running travel speeds combined) is provided in Figure~\ref{fig:speed_cor_total} Left. In this case, a power function was added as a fitting line for illustrative purposes due to a more appropriate curve shape. The results for travel speeds in corridors in different density intervals are tabulated in \ref{app:cor} in Table~\ref{tab:speed_cor} (walking and running travel speeds combined), Table~\ref{tab:speed_cor_walk} (walking speeds only) and Table~\ref{tab:speed_cor_run} (running speeds only).
The observed travel speeds changed with the ages of the children; notably, they increased with increasing age. On average, the walking Junior children traveled at speeds of 1.06~m$\cdot$s$^{-1}$; Senior children walked 1.38~m$\cdot$s$^{-1}$; and Senior+ children 1.92~m$\cdot$s$^{-1}$ in the same density interval. The maximum values of the measured walking travel speeds were relatively high, corresponding to the fast walking (\quotes{race-walking}) gait of some children instructed to walk and not run. Maximum running speeds for Junior children were less than the maximum walking speeds observed for Senior children. Differences between age groups were also found when comparing the frequency of walking and running by children. The results provided in Table~\ref{tab:walk_run} suggest that older children (Senior and especially Senior+ children) appeared to run more frequently than Junior children. This may likely be attributed to the children's different proficiency in running and/or to the distinctive instructions and levels of supervision provided by staff to children in different age groups. \par
\begin{table}[hbt!]
    \centering
    \footnotesize
        \begin{tabular}{R{1.6cm}|R{5cm}|R{5cm}}
        \hline
         \textbf{Age group} & \textbf{Frequency of walking children [\%] (data points)} & \textbf{Frequency of running children [\%] (data points)} \\
         \hline
         Junior & 74.7 (139) & 25.3 (47) \\
         \hline
         Senior & 59.1 (153) & 40.9 (104) \\
         \hline
         Senior+ & 17.0 (15) & 83.0 (73) \\
         \hline
         Mixed & 28.4 (50) & 71.6 (126) \\
         \hline
         \multicolumn{1}{l|}{\textbf{Total}} & \multicolumn{1}{l|}{\textbf{50.4 (357)}} & \multicolumn{1}{l}{\textbf{49.6 (350)}} \\
         \hline
             \end{tabular}
    \caption{Frequency of children who were walking and running during the evacuation drills.}
    \label{tab:walk_run}
\end{table}
The dependence of specific flow $F_s$ on density $D$ for different age groups (color coded) is demonstrated in Figure~\ref{fig:speed_cor_total} Right (tabulated results can be found in \ref{app:cor} in Table~\ref{tab:flow_cor}). The specific flows observed in the corridors ranged from 0.65~pers$\cdot$s$^{-1}$\,m$^{-1}$ to 4.17~pers$\cdot$s$^{-1}$\,m$^{-1}$ (mean value of 1.78~pers$\cdot$s$^{-1}$\,m$^{-1}$). In contrast to the observed results for travel speeds, children's ages did not appear to impact specific flows in corridors. For the continuous movement of children (686~data points), densities ranged from 0.00~to 0.05~m$^2$\,m$^{-2}$ (i.e.,~to approx. 1.9~children$\cdot$m$^{-2}$). Higher density conditions between 0.07~and 0.09~m$^2$\,m$^{-2}$ were registered only for the case where one Senior class stopped inside a measurement area (see the data points indicated with cross symbols in Figures~\ref{fig:speed_cor_walk+run} and ~\ref{fig:speed_cor_total}). Therefore, due to the limited number of observations under higher density conditions, the application range of the given trend lines is recommended for densities lower than 0.05~m$^2$\,m$^{-2}$ (approximately 1.9~children$\cdot$m$^{-2}$; see the solid parts of the trend lines in Figures~\ref{fig:speed_cor_walk+run} and ~\ref{fig:speed_cor_total}). The low densities measured can be attributed to the free movement conditions in the corridors and the relatively high travel speeds reached by the observed children, especially those who ran. 

\subsection{Movement on straight staircases}
Travel speeds, pedestrian flows, and densities were analyzed separately for flights (27~measurement areas), landings (11~measurement areas), and entire staircases (i.e., for flights and landings together; stairs with only one flight were not included in this category; nine measurement areas in total) observing 13~straight staircases in six nursery schools during seven evacuation drills. All staircases observed were parts of participants' daily circulation routes and well known to them (further details on the observed staircases including their geometries can be found in \cite{najmanova_evacuation_2020}).\par 

\subsubsection{Flights of straight staircases}
\begin{figure}
    \centering
    \includegraphics[width=\textwidth]{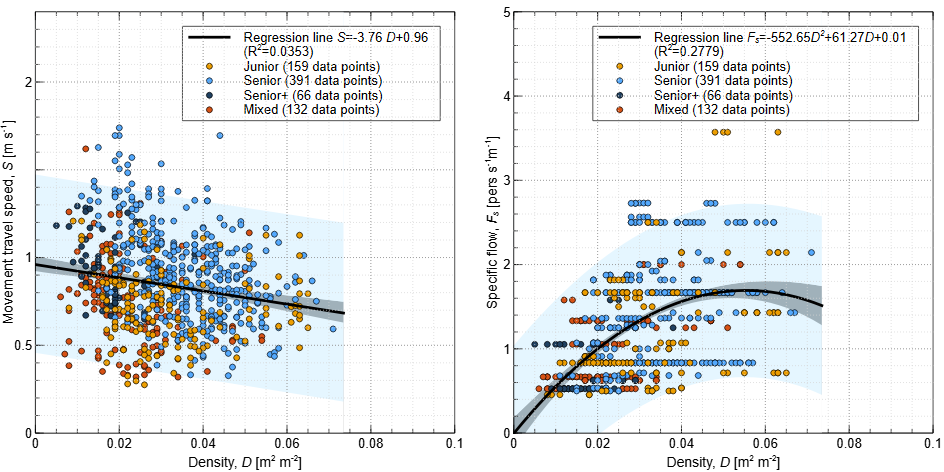}
    \vspace{-8mm}
    \caption{
    \textbf{Left}: Travel speeds (walking and running combined) per density on flights of straight staircases;  
    \textbf{Right}: Specific flows per density on flights of straight staircases.
    \textbf{Note}: Regression line, 95\% confidence bounds of the fitted curve (grey area), and 95\% prediction bounds for a future observation (blue area) are included.}
            \vspace{-8mm}
    \label{fig:speed_stair_fl_total}\
\end{figure}

On flights of straight staircases, walking and running travel speeds were not distinguished because it was not possible to identify the flight phase in the children's movement. Figure~\ref{fig:speed_stair_fl_total} Left presents the dependence of movement travel speed $S$ on density $D$ for different age groups (color coded). The presented data set is scattered, without any clear observable trends, which may be attributed to the large diversity of the staircases observed as well as to the variability of pedestrian dynamics. The results for travel speeds on flights in different density intervals are tabulated in \ref{app:stair} in Table~\ref{tab:speed_stair_fl}. 
Observations showed that the travel speeds of the children travelling on flights were age-dependent. Junior children moved at lower speeds (mean value 0.71~m$\cdot$s$^{-1}$ in a density interval 0.00-0.05~m$^2$\,m$^{-2}$) than Senior and Senior+ children (mean value 0.94~m$\cdot$s$^{-1}$ in the same density interval). However, age differences for travel speeds on flights were not as significant as they were in corridors where older children's speeds were almost twice as high as those observed for younger children. This may be attributed to the fact that moving downstairs may be a more challenging task for younger children. Consequently, the travel speeds observed on flights were considerably lower than those measured in corridors. Due to the free and continuous movement of children on flights without stops or crowding, only low density conditions (less than 0.1~m$^2$\,m$^{-2}$) occurred on flights during the evacuation drills. \par  
%
The dependence of specific flow $F_s$ on density $D$ is illustrated in Figure~\ref{fig:speed_stair_fl_total} Right. The observed values for specific flow on flights were slightly lower than the results obtained for the corridors, ranging from 0.45~pers$\cdot$s$^{-1}$\,m$^{-1}$ to 3.57~pers$\cdot$s$^{-1}$\,m$^{-1}$ (mean value 1.28~pers$\cdot$s$^{-1}$\,m$^{-1}$). Following the estimated trend line, the peak of specific flow (1.7~pers$\cdot$s$^{-1}$\,m$^{-1}$) can be identified for a density of around 0.055~m$^2$\,m$^{-2}$, i.e.,~approximately 2~children$\cdot$m$^{-2}$. 
The results obtained for specific flows at different density intervals are summarised in \ref{app:stair} in Table~\ref{tab:flow_stair_fl}. Due to the absence of experimental data at higher densities, the application range of the provided data sets should be considered limited to densities lower than 0.07~m$^2$\,m$^{-2}$ (approximately 2.7~children$\cdot$m$^{-2}$).

\subsubsection{Landings of straight staircases}
\begin{figure}[hbt!]
    \centering
    \includegraphics[width=\textwidth]{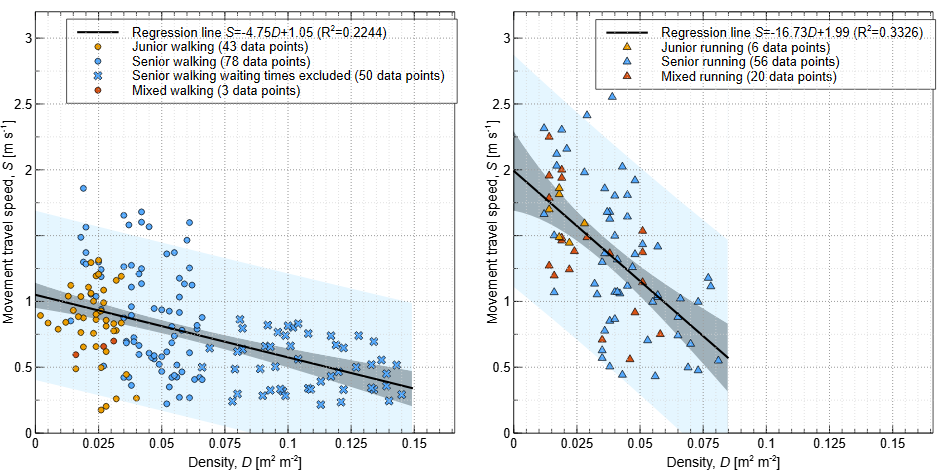} 
        \vspace{-8mm}
        \caption{
        \textbf{Left}: Walking travel speeds per density on landings of straight staircases for different age groups;
        \textbf{Right}: Running travel speeds per density on landings of straight staircases for different age groups.
    \textbf{Note}: Regression line, 95\% confidence bounds of the fitted curve (grey area), and 95\% prediction bounds for a future observation (blue area) are included.}
    \label{fig:speed_stair_land_group}
\end{figure}

\begin{figure}
    \centering
    \includegraphics[width=\textwidth]{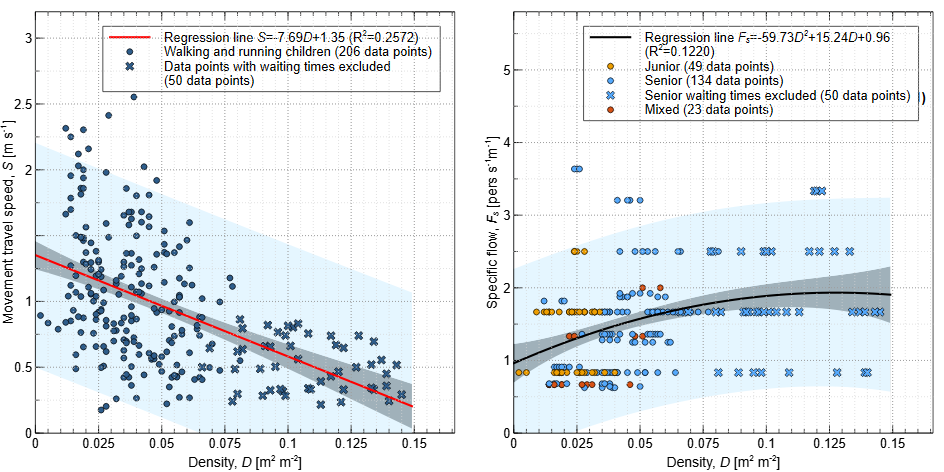}
    \vspace{-8mm}
    \caption{
    \textbf{Left}: Travel speeds (walking and running combined) per density on landings of straight staircases;  
    \textbf{Right}: Specific flows per density on landings of straight staircases.
    \textbf{Note}: Regression line, 95\% confidence bounds of the fitted curve (grey area), and 95\% prediction bounds for a future observation (blue area) are included.}
    \label{fig:speed_stair_land_total}\
        \vspace{-8mm}
\end{figure}

Travel speeds on landings were measured during continuous movement (i.e., movement travel speeds) except for two Senior classes (35~children, 50~data points) whose waiting times in the measurement areas were excluded to calculate their travel speeds. 
In Figure~\ref{fig:speed_stair_land_group}, the dependence of movement travel speed $S$ on density $D$ is provided separately for walking and running children for different age groups (color coded). The speed-density relationship for all observations (all age groups, walking and running travel speeds combined) can be seen in Figure~\ref{fig:speed_stair_land_total} Left. The results for travel speeds on landings of straight staircases at different density intervals are given in \ref{app:stair} in Table~\ref{tab:speed_stair_land} (walking and running travel speeds combined), Table~\ref{tab:speed_stair_str_walk} (walking speeds only) and Table~\ref{tab:speed_stair_str_run} (running speeds only).\par
The results revealed that the children moved faster on landings (average value for all children 0.96~m$\cdot$s$^{-1}$) than on flights of straight staircases (average value for all children was 0.85~m$\cdot$s$^{-1}$). Similarly, the maximum travel speeds reached by children of various age were higher on landings (2.55~m$\cdot$s$^{-1}$) than on flights (1.74~m$\cdot$s$^{-1}$). This appears to correspond to findings indicating that horizontal planes enable children to move at higher speeds. However, the walking and running travel speeds observed on landings were considerably lower than the travel speeds in corridors. This may be because the landings have limited space for movement and/or the landing shapes play a role in movement trajectories. Age-dependence of travel speed can be identified by comparing the results for Junior and Senior children, especially for walking travel speeds. Differences in maximum running speeds were also observed for Junior children (1.86~m$\cdot$s$^{-1}$; 0.00-0.05~m$^2$\,m$^{-2}$) and Senior children (2.55~m$\cdot$s$^{-1}$ in the same density interval). Compared to movement on flights and in corridors, higher densities (over 0.1~m$^2$\,m$^{-2}$ with a maximum value of 0.14~m$^2$\,m$^{-2}$, i.e.,~approximately 5.2~children$\cdot$m$^{-2}$) were observed on landings. These higher density conditions appear to have occurred when 35~Senior children had to stop inside a measurement area, are clearly visible in Figure~\ref{fig:speed_stair_land_total} (see the cross symbols). \par
The dependence of specific flow $F_s$ on density $D$ is demonstrated in Figure~\ref{fig:speed_stair_land_total} Right (tabulated results are summarised in \ref{app:stair} in Table~\ref{tab:flow_stair_land}). Because the limited number of data points and data analysis methods employed resulted in discrete jumps in the scattered plot, the fitting curve has an indicative purpose in this graph. The peak of the suggested trend line (close to 2.0~pers$\cdot$s$^{-1}$\,m$^{-1}$) can be estimated at a density 0.13~m$^2$\,m$^{-2}$, i.e.,~approximately 5~children$\cdot$m$^{-2}$. The results for specific flow on landings ranged from 0.63~pers$\cdot$s$^{-1}$\,m$^{-1}$ to 3.64 pers$\cdot$s$^{-1}$\,m$^{-1}$ (mean value 1.52~pers$\cdot$s$^{-1}$\,m$^{-1}$). Compared to observations for flights, higher specific flow values were measured on landings. Although lower, these results were closer to the specific flow values recorded in the corridors. The application range of the trend lines describing the movement on landings is limited to densities lower than 0.14~m$^2$\,m$^{-2}$ (approximately 5~children$\cdot$m$^{-2}$).
\subsubsection{Entire straight staircases}
\begin{figure}
    \centering
    \includegraphics[width=\textwidth]{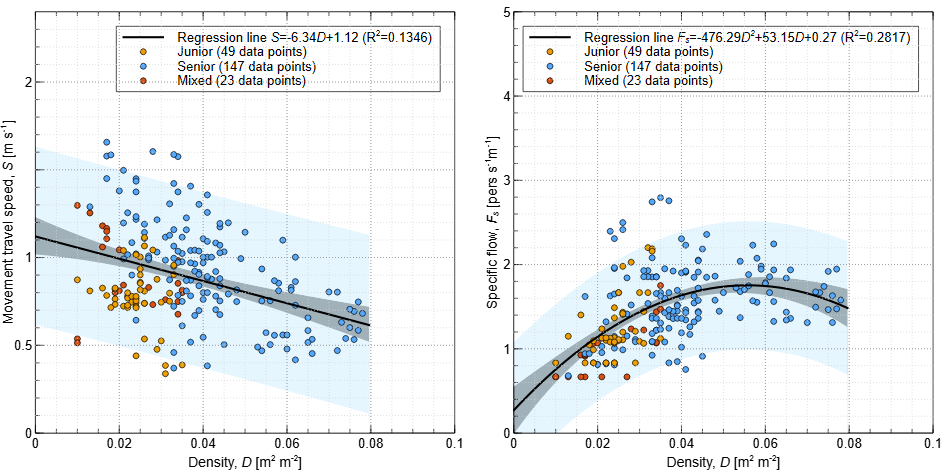}
    \vspace{-8mm}
    \caption{
    \textbf{Left}: Travel speeds (walking and running combined) per density on entire straight staircases;  
    \textbf{Right}: Specific flows per density on entire staircases.
    \textbf{Note}: Regression line, 95\% confidence bounds of the fitted curve (grey area), and 95\% prediction bounds for a future observation (blue area) are included.}
        \vspace{-8mm}
    \label{fig:speed_stair_whole_total}\
\end{figure}

Because children's movement on flights and landings was evaluated as continuous movement (or modified to be assumed as continuous in some cases), entire straight staircases were also observed to analyse movement travel speeds on them. The relationships between travel speeds and densities estimated for entire straight staircases are illustrated in Figure~\ref{fig:speed_stair_whole_total} Left (age groups are color coded). The results follow the previously described trends observed for flights and landings (notably, age and density appear to play a role in travel speeds). Because children spent more time moving on flights than on landings, the results provided for entire staircases appear to have been impacted by their overall movement patterns. The results for travel speeds at different density intervals are provided in \ref{app:stair} in Table~\ref{tab:speed_stair_whole}. \par
The flow-density relationships estimated for different age groups (color coded) are demonstrated in Figure~\ref{fig:speed_stair_whole_total} Right (tabulated results for different density intervals are provided in \ref{app:stair} in Table~\ref{tab:flow_stair_whole}). According to the trend lines provided, the peak of specific flow (almost 1.8~pers$\cdot$s$^{-1}$\,m$^{-1}$) occurred at a density 0.05~m$^2$\,m$^{-2}$, i.e.,~approximately 2~children$\cdot$m$^{-2}$. The application range of the trend lines provided for entire straight staircases is defined for densities lower than 0.08~m$^2$\,m$^{-2}$ (approximately 3.1~children$\cdot$m$^{-2}$).   
\subsection{Movement on spiral staircases}
\begin{figure}[hbt!]
    \centering
    \includegraphics[width=\textwidth]{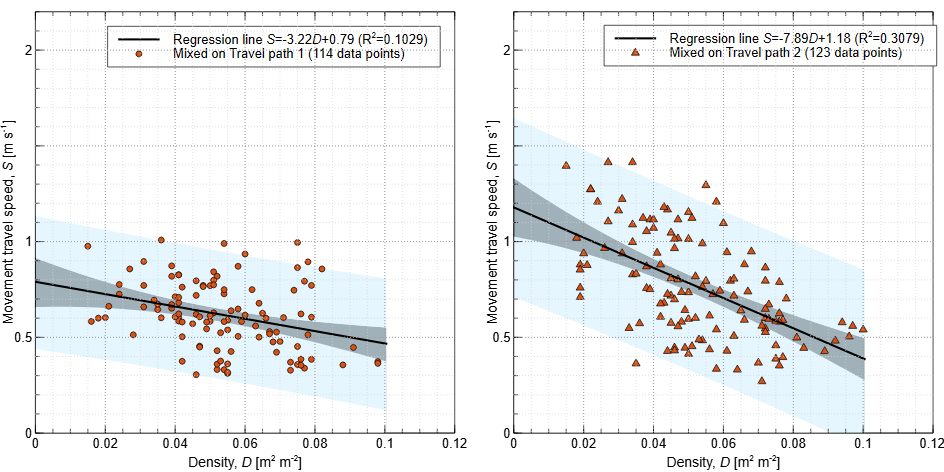} 
        \vspace{-8mm}
    \caption{
    \textbf{Left}: Travel speeds per density on spiral staircases on Travel path~1;  
    \textbf{Right}: Travel speeds per density on spiral staircases on Travel path~2.
    \textbf{Note}: Regression line, 95\% confidence bounds of the fitted curve (grey area), and 95\% prediction bounds for a future observation (blue area) are included.}
    \label{fig:speed_stair_spiral_path}
\end{figure}

\begin{figure}
    \centering
    \includegraphics[width=\textwidth]{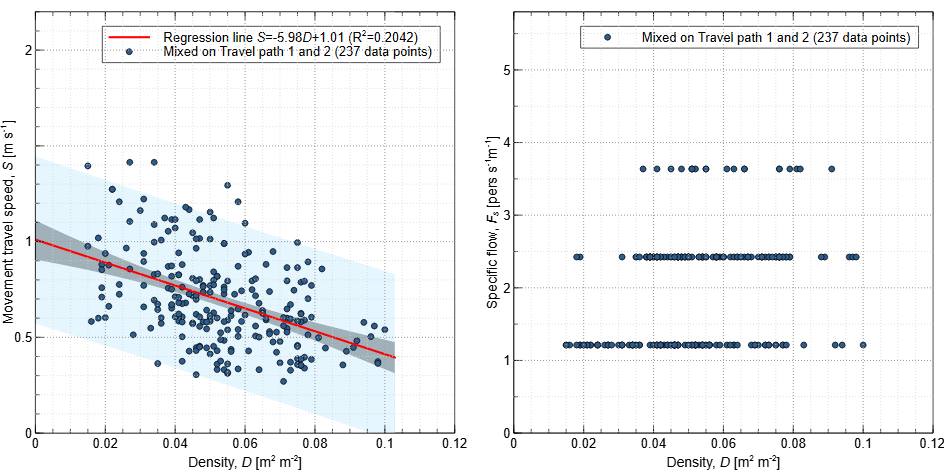}
        \vspace{-8mm}
    \caption{
    \textbf{Left}: Travel speeds per density on spiral staircases (Travel path~1 and~2 combined);  \textbf{Right}: Specific flows per density on spiral staircases.
    \textbf{Note}: Regression line, 95\% confidence bounds of the fitted curve (grey area), and 95\% prediction bounds for a future observation (blue area) are included for speed-density observations.}
    \label{fig:speed_stair_spiral_total}\
\end{figure}

Travel speeds, pedestrian flows, and densities were observed on two identical spiral staircases (eight measurement areas) located outside one nursery school during one evacuation drill (only Mixed classes involved). To explore the potential impact of the different geometry of stairs, travel speeds for children moving on Travel path~1 and Travel path~2 (see Figure~\ref{fig:spiral}) were analyzed separately (the difference in slope between the assumed travel paths was almost 10$^{\circ}$; details for the geometry of the staircases can be found in \cite{najmanova_evacuation_2020}). Walking and running travel speeds were not distinguished. The speed-density relationships considering different travel paths are presented in Figure~\ref{fig:speed_stair_spiral_path}, the speed-density relationship for all observations (Travel paths~1 and~2 combined) can be seen in Figure~\ref{fig:speed_stair_spiral_total} Left (see also Tables~\ref{tab:speed_stair_spiral_1} and~\ref{tab:speed_stair_spiral_2}, resp. Table~\ref{tab:speed_stair_spiral} in \ref{app:stair_spiral}, for tabulated results).
The observed travel speeds on external spiral staircases ranged from 0.27~m$\cdot$s$^{-1}$ to 1.41~m$\cdot$s$^{-1}$ (mean value of 0.69~m$\cdot$s$^{-1}$). Comparing these results with the data sets describing travel speeds for Mixed children on flights of internal straight staircases, the children's travel speeds were similar (see Table~\ref{tab:speed_stair_fl}). Lower travel speeds were observed for children moving on Travel path~1 (mean value 0.61~m$\cdot$s$^{-1}$]) where the movement space was more limited than on Travel path~2 (mean value 0.76~m$\cdot$s$^{-1}$) in all density intervals; however, greater differences in travel speeds were observed at lower densities (0.00-0.05~m$^2$\,m$^{-2}$). In addition, when children moved in pairs (i.e., in two lanes), those walking downstairs on Travel path~2 had to slow down to adjust their travel speeds for slower partners. Hence, the impact of the geometries of spiral stairs may be considered even more significant than evident from the measured values. The application range of the provided trend lines should be considered limited to density conditions up to 0.1~m$^2$\,m$^{-2}$ (approximately 3.8~children$\cdot$m$^{-2}$). \par
Due to the limited number of data points measured on identical staircases with the same assumed widths and data analysis methods used, the effect of discrete jumps resulted in only three values of specific flow (see Figure~\ref{fig:speed_stair_spiral_total} Right). Hence, the scatter plot was included only for informative purposes without any regression analysis of the data set. Compared to the specific flows measured for straight staircases, higher specific flow values were observed on spiral staircases, which may have been related to more crowded conditions for children going downstairs side-by-side. The results for specific flows at different density intervals are provided in \ref{app:stair_spiral} in Table~\ref{tab:flow_stair_spiral}. 

\subsection{Movement through doorways}
\begin{figure}[hbt!]
    \centering
    \includegraphics[width=\textwidth]{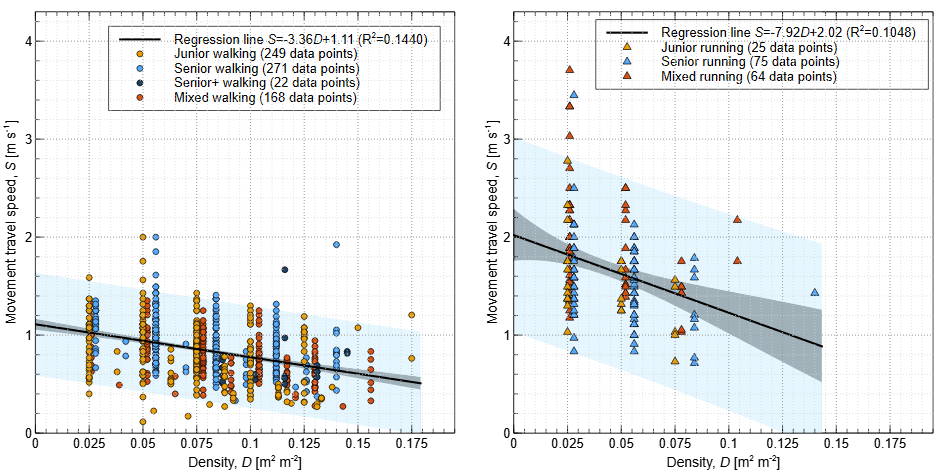} 
        \vspace{-8mm}
        \caption{
        \textbf{Left}: Walking travel speeds per density in doorways for different age groups;
        \textbf{Right}: Running travel speeds per density in doorways for different age groups.
    \textbf{Note}: Regression line, 95\% confidence bounds of the fitted curve (grey area), and 95\% prediction bounds for a future observation (blue area) are included.}
    \label{fig:speed_door_walk+run}
\end{figure}

\begin{figure}
    \centering
    \includegraphics[width=\textwidth]{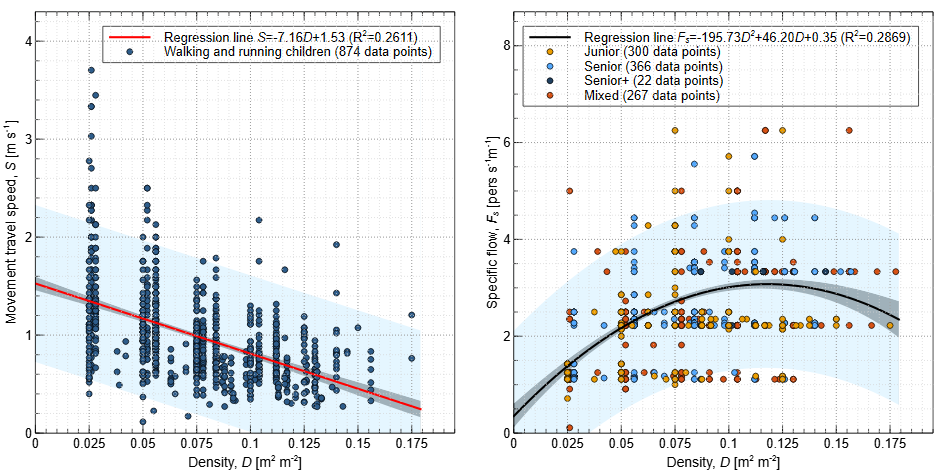}
    \vspace{-8mm}
    \caption{
    \textbf{Left}: Travel speeds (walking and running combined) per density in doorways;  
    \textbf{Right}: Specific flows per density in doorways.
    \textbf{Note}: Regression line, 95\% confidence bounds of the fitted curve (grey area), and 95\% prediction bounds for a future observation (blue area) are included.}
    \label{fig:speed_door_total}\
        \vspace{-8mm}
\end{figure}

Travel speeds, pedestrian flows, and densities were measured in 25~measurement areas in nine participating nursery schools during 14~experimental evacuation drills. The width of the observed doors was between 0.7~to 1.4~m (average value of 0.88~m). In Figure~\ref{fig:speed_door_walk+run}, movement travel speed $S$ through doorways is represented as dependent on density $D$ separately for children walking or running in different age groups (color coded; see Tables~\ref{tab:speed_door_walk} and~\ref{tab:speed_door_run} for tabulated results). The speed-density relationship for all observations (walking and running children combined) is presented in Figure~\ref{fig:speed_door_total} Left (tabulated results are provided in \ref{app:door} in Table~\ref{tab:speed_door}). Due to the fixed size of the measurement areas in doorways and the assumption that individual participants were discrete units, jumps representing discrete density values can be identified in the plots.  
Comparing the walking travel speeds observed in doorways with the corresponding results measured in other horizontal evacuation routes locations (i.e.,~in corridors and on landings of straight staircases), the values measured in doorways were similar to the observations for straight staircases landings and were therefore considerably lower than those in corridors. This may be related to the specific conditions for movement in the doorway measurement areas, more restricted by a bottleneck, doors, than corridors. Consequently, densities up to 0.18~m$^2$\,m$^{-2}$ (i.e.,~approximately 7~children$\cdot$m$^{-2}$) occurred in doorways, which were almost double the values observed for corridors or staircases (less than 0.1~m$^2$\,m$^{-2}$; except for the situations when children stopped on landings, when the observed density reached 0.14~m$^2$\,m$^{-2}$). In contrast, higher travel (mostly running) speeds were observed in doorways at lower densities (without bottlenecks), when the conditions for participants' movement through doorways were similar to movement in corridors. \par
The flow-density relationship related to children's movement in doorways is shown in Figure~\ref{fig:speed_door_total} Right (tabulated results can be found in \ref{app:door} in Table~\ref{tab:flow_door}). Following the given trend lines, the peak that describes the maximum specific flow (approximately 3~pers$\cdot$s$^{-1}$\,m$^{-1}$) can be identified at density of around 0.12~m$^2$\,m$^{-2}$, i.e.,~approximately 4.6~children$\cdot$m$^{-2}$). Considering all observations, the observed specific flow in doorways ranged from 0.11~pers$\cdot$s$^{-1}$\,m$^{-1}$ to 6.25~pers$\cdot$s$^{-1}$\,m$^{-1}$ with a mean value of 2.51~pers$\cdot$s$^{-1}$\,m$^{-1}$. Higher values of specific flow were observed for Senior children compared to Junior children for the same density levels. Compared to the measurements performed in corridors and on staircases, the results obtained for doorways represent the highest values of specific flows in this study due to crowded conditions at the doors (bottlenecks). The application range of the data sets describing the movement through doorways is limited to densities lower than 0.18~m$^2$\,m$^{-2}$, i.e.,~approximately 7~children$\cdot$m$^{-2}$. \par
\subsection{Movement behaviour}
The need for help from staff members and children holding hands was observed in all 100~measurement areas and 15~evacuation drills. On staircases, the use of handrails and a \quotes{marking time} pattern by children was observed in 46~measurement areas. Because children appeared to change their behaviours within the measurement areas, their movement behaviour was monitored separately in the first and second halves of each measurement area; except for doorways, due to a small size of measurement areas at doors.   
\subsubsection{Levels of physical assistance provided to children}
Physical support provided by staff members to children was described using five categories. The category \quotes{No physical assistance} includes all cases in which no physical contact was observed between children and staff members. The category \quotes{Gentle pushing} typically indicates a gentle push on a child's back, shoulders, arms, or head. This contact was not necessary to proceed the evacuation process and did not represent physical support. Physical contacts necessary to facilitate evacuation process (e.g., redirecting and pushing children in the right direction, speeding them up, keeping them moving) were classified as \quotes{Physical contact needed}. Cases in which staff members held children's hands and carried them were designated as separate categories (\quotes{Hand holding} and \quotes{Carried}). 
\begin{table}[hbt!]
    \centering
    \footnotesize
    \begin{tabular}{R{3.7cm}|R{2.0cm}|R{1.7cm}|R{1.7cm}|R{1.7cm}|R{1.7cm}}
    \hline
    \multirow{2}{=}{\textbf{Physical assistance provided}} & \multicolumn{5}{R{7.5cm}}{\textbf{Frequency [\%] (data points)}} \\
    \cline{2-6}
    & \textbf{Total} & \textbf{Junior} & \textbf{Senior} & \textbf{Senior+} & \textbf{Mixed} \\
    \hline
    No physical assistance & \textbf{90.1 (3640)} & 88.5 (862) & 92.1 (1567) & 97.0 (256) & 86.7 (955) \\
    \hline
    Gentle pushing & \textbf{4.4 (176)} & 3.4 (33) & 5.0 (85) & 0.8 (2) & 5.1 (56) \\
    \hline
    Physical contact needed & \textbf{1.2 (48)} & 2.2 (21) & 0.1 (2) & 0.0 (0) & 2.3 (25)  \\
    \hline
    Hand holding & \textbf{4.3 (174}) & 6.0 (58) & 2.8 (47) & 2.3 (6) & 5.7 (63) \\
    \hline
    Carried & \textbf{0.1 (3)} & 0.0 (0) & 0.0 (0) & 0.0 (0) & 0.3 (3) \\
    \hline
    \end{tabular}
    \caption{Levels of physical assistance provided to children in different age groups for all parts of evacuation routes during evacuation drills.}
    \label{tab:mov_contact_all}
\end{table}
The results obtained for different age groups in all evacuation routes in total (corridors, straight and spiral staircases, doorways) are summarised in Table~\ref{tab:mov_contact_all}. Separate results for different parts of the evacuation routes can be found in \ref{app:beh} in Table~\ref{tab:mov_contact}. 
In 90.1\% of the observations, no physical contact was observed between children and staff members. Physical contact of some sort was identified more often in corridors and doorways than on staircases (92.2\% of children on straight staircases, 97.6\% on spiral staircases did not receive any assistance). This resulted from different types of supervision. In corridors and doorways, staff members often stopped their movements and occasionally touched passing children. In contrast, such behaviours were not observed on staircases where staff members were busy helping slower or less experienced children. In addition, no \quotes{Gentle pushing} contacts were recorded on spiral staircases due to space limitations. 
Junior children and younger children in Mixed classes received more physical assistance, especially \quotes{Physical contact needed} and \quotes{Hand holding}. In contrast, \quotes{Gentle pushing} was the most often seen contact for Senior children. This suggests that staff members' approaches and behaviour towards children were to a large extent influenced by the age of the children. \par
\subsubsection{Hand holding}
To investigate hand holding during evacuation drills, four categories were determined: \quotes{No hand holding}, \quotes{With another child}, \quotes{With a staff member}, and \quotes{With another child and a staff member}. The results for different age groups for all parts of the evacuation routes (corridors, straight and spiral staircases, doorways) are provided in Table~\ref{tab:mov_hand_all}. The results for different parts of the evacuation routes are given separately in \ref{app:beh} in Table~\ref{tab:mov_hand}.
\begin{table}[hbt!]
    \centering
    \footnotesize
    \begin{tabular}{R{3.8cm}|R{2cm}|R{1.7cm}|R{1.7cm}|R{1.7cm}|R{1.6cm}}
    \hline
    \multirow{2}{=}{\textbf{Hand holding}} & \multicolumn{5}{R{7.5cm}}{\textbf{Frequency [\%] (data points)}} \\
    \cline{2-6}
    & \textbf{Total} & \textbf{Junior} & \textbf{Senior} & \textbf{Senior+} & \textbf{Mixed} \\
    \hline
     No hand holding & \textbf{59.2 (2392)} & 50.1 (488) & 43.5 (740) & 81.8 (214) & 86.2 (950) \\
    \hline
    With another child & \textbf{36.5 (1475)} & 43.9 (428) & 53.7 (914) & 16.7 (44) & 8.1 (89) \\
    \hline
    With a staff member & \textbf{3.9 (156)} & 4.4 (43) & 2.6 (44) & 2.3 (6) & 5.7 (63)  \\
    \hline
    With another child and a staff member & \textbf{0.4 (18)} & 1.5 (15) & 0.2 (3) & 0.0 (0) & 0.0 (0) \\
    \hline
  \end{tabular}
\caption{Hand holding observed for different age groups and in all parts of the evacuation routes during evacuation drills.}
\label{tab:mov_hand_all}
\end{table}
The children moved individually without holding hands in 59.2\% and formed pairs in 36.5\% of all observations. A higher frequency of hand holding was observed in corridors and doorways than in staircases. This may be attributed to familiarity with daily routines or more difficult conditions making moving in pairs uncomfortable on stairs. On wider flights, staff members often instructed children, especially Junior children, to hold the handrail and not hands. This contributed, for straight staircases, to almost half the hand holding levels observed in corridors. On spiral staircases, hand holding limited even more because of less wide and steeper spaces.
In schools with small capacities and simple layouts, children tended to leave the building individually (i.e., not in pairs) more often than in schools with large capacities, where greater evacuation organization was needed. The impact of children's age is evident in the categories \quotes{With a staff member} and \quotes{With another child and a staff member}, because staff members provided more guidance to younger children. 
\subsubsection{Use of handrails}
For analysis of the use of handrails on staircases, handrails at standard heights (\quotes{(standard handrail)}, lower heights (\quotes{children's handrail}), and \quotes{Baluster or other support} (e.g.,~walls) were observed. A summary of the observations made for straight staircases is provided in Table~\ref{tab:mov_handrails_all}; the results obtained for spiral staircases are given in Table~\ref{tab:mov_handrails_spiral}. Since different combinations of handrail types were available in nursery schools, the results for situations where only standard handrails, only children's handrails, and both standard and children's handrails were present are shown separately in \ref{app:beh} in Table~\ref{tab:mov_handrails}.
\begin{table}[hbt!]
    \centering
    \footnotesize
    \begin{tabular}{R{3.9cm}|R{1.8cm}|R{1.7cm}|R{1.7cm}|R{1.7cm}|R{1.7cm}}
    \hline
    \multirow{2}{=}{\textbf{Use of handrails on straight staircases}} & \multicolumn{5}{R{7.5cm}}{\textbf{Frequency [\%] (data points)}} \\
    \cline{2-6}
    & \textbf{Total} & \textbf{Junior} & \textbf{Senior} & \textbf{Senior+} & \textbf{Mixed} \\
    \hline
    Without the use of handrails & \textbf{32.4 (463)} & 28.4 (78) & 36.7 (293) & 2.7 (3) & 36.6 (89) \\
    \hline
    Standard handrail & \textbf{43.2 (616)} & 53.1 (146) & 32.4 (259) & 97.3 (107) & 42.8 (104) \\
    \hline    
    Children's handrail & \textbf{22.0 (314)} & 14.5 (40) & 28.0 (224) & 0.0 (0) & 20.6 (50) \\
    \hline
    Baluster or other support & \textbf{2.4 (34)} & 4.0 (11) & 2.9 (23) & 0.0 (0) & 0.0 (0) \\
    \hline
  \end{tabular}
   \caption{Handrail use observed for different age groups on straight staircases during evacuation drills}
    \label{tab:mov_handrails_all}
\end{table}
\begin{table}[hbt!]
    \centering
    \footnotesize
    \begin{tabular}{R{4cm}|R{1.7cm}|R{1.7cm}|R{1.7cm}|R{1.7cm}|R{1.7cm}}
    \hline
    \multirow{2}{=}{\textbf{Use of handrails on spiral staircases}} & \multicolumn{5}{R{7.5cm}}{\textbf{Frequency [\%] (data points)}} \\
    \cline{2-6}
    & \textbf{Total} & \textbf{Junior} & \textbf{Senior} & \textbf{Senior+} & \textbf{Mixed} \\
\hline
    \multicolumn{6}{l}{\textbf{Both standard and children's handrail available}} \\
    \hline
    Without the use of handrails & 13.4 (44) & N/A & N/A & N/A & 13.4 (44) \\
    \hline
    Standard handrail & 42.2 (139) & N/A & N/A & N/A & 42.2 (139) \\
    \hline
    Children's handrail & 30.7 (101) & N/A & N/A & N/A & 30.7 (101)  \\
    \hline
    Baluster or other support & 13.7 (45) & N/A & N/A & N/A & 13.7 (45) \\
    \hline
  \end{tabular}
    \caption{Handrail use observed for different age groups on spiral staircases during evacuation drills}
    \label{tab:mov_handrails_spiral}
\end{table}
The children did not use any handrail in 32.4\% of observations, whereas they used standard handrails in 43.2\% and children's handrails in 22.0\%  or another support (baluster or wall) in 2.4\%, respectively. 
Where both standard and children's handrails were present, the children used them at approximately the same levels (33.1\% standard handrail, 29.8\% children's handrail). Children's handrails were more frequently used by Mixed children and standard handrails by Junior children; Senior children used both types almost equally. 
The general assumption that Junior children would prefer lower handrails was not confirmed. The decision to use standard handrails may perhaps be explained as a demonstration by children of their ability to do what adults do. However, Junior children used handrails in 71.6\% of the observations, the highest value of all age groups. As with hand holding, the use of handrails by children was largely impacted by daily routines and instructions given by staff members. On spiral staircases, children used a support in 86.8\% of observations. Due to the round shape of such staircases, children used balusters (vertical elements) more frequently (13.7\%) than in straight staircases.  \par
\subsubsection{Marking time patterns}
\quotes{Marking time} refers to an action during which one foot meets the other on a step (i.e.,~both feet are on the same level), before going to the next higher (or lower) step \cite{williams_stair_1994}. The results summarised in Table~\ref{tab:mov_marking} show that marking time patterns were observed more frequently for Junior children and Mixed children (especially younger children in a group). This finding is in line with the literature suggesting such an alternating pattern emerges until four years of age \cite{gesell_first_1940, berk_child_2006}. The marking time pattern was observed to be the result of children's limited movement abilities, insufficient space available for alternating feet (i.e., the step in front of a child was occupied), or from a play. \par 
\begin{table}[hbt!]
    \centering
    \footnotesize
    \begin{tabular}{R{4cm}|R{1.7cm}|R{1.7cm}|R{1.7cm}|R{1.7cm}|R{1.7cm}}
    \hline
    \multirow{2}{=}{\textbf{Marking time pattern}} & \multicolumn{5}{R{7.5cm}}{\textbf{Frequency [\%] (data points)}} \\
    \cline{2-6}
    & \textbf{Total} & \textbf{Junior} & \textbf{Senior} & \textbf{Senior+} & \textbf{Mixed} \\
\hline
    \multicolumn{6}{l}{\textbf{Straight staircases}} \\
    \hline
    Used & 9.8 (150) & 19.8 (68) & 4.5 (38) & 0.0 (0) & 18.1 (44) \\
    \hline
    Not used & 90.2 (1383) & 80.2 (275) & 95.5 (799) & 100.0 (110) & 81.9 (199) \\
\hline
    \multicolumn{6}{l}{\textbf{Spiral staircases}} \\
    \hline
    Used & 18.8 (62) & N/A & N/A & N/A & 18.8 (62) \\
    \hline
    Not used & 81.2 (267) & N/A & N/A & N/A & 81.2 (267) \\
    \hline
  \end{tabular}
    \caption{Use of marking time pattern observed for different age groups on staircases during evacuation drills}
    \label{tab:mov_marking}
\end{table}

\section{Discussion}\label{sec:discuss}
This section compares the results presented with relevant research studies focused on the evacuation movement and provides a new set of behavioural statements on the evacuation behaviour of preschool children in nursery schools.
\subsection{Comparison with previous research studies}\label{subsec:comparison}
\begin{figure} [hbt!]
    \centering
    \includegraphics[width=\textwidth] {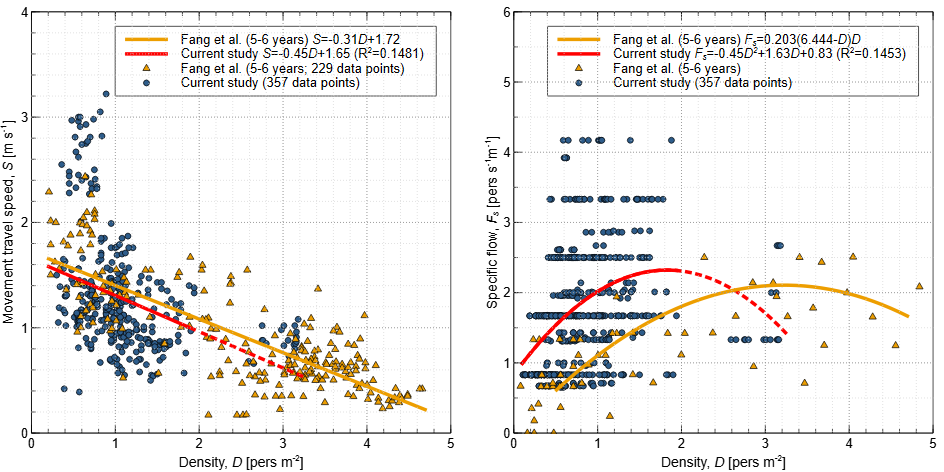}
        \vspace{-8mm}
        \caption{Movement characteristics of preschool children observed in corridors (walking children) compared to Fang et al.~\cite{fang_experimental_2019}.  
        \textbf{Left}: Walking travel speeds;   
        \textbf{Right}: Specific flows.
        \textbf{Note}: Dashed parts of the trend lines indicate density intervals for a limited number of data points.} 
    \label{fig:speed_cor_comparII}
\end{figure}

\begin{figure}[hbt!]
    \centering
    \includegraphics[width=1.0\textwidth]{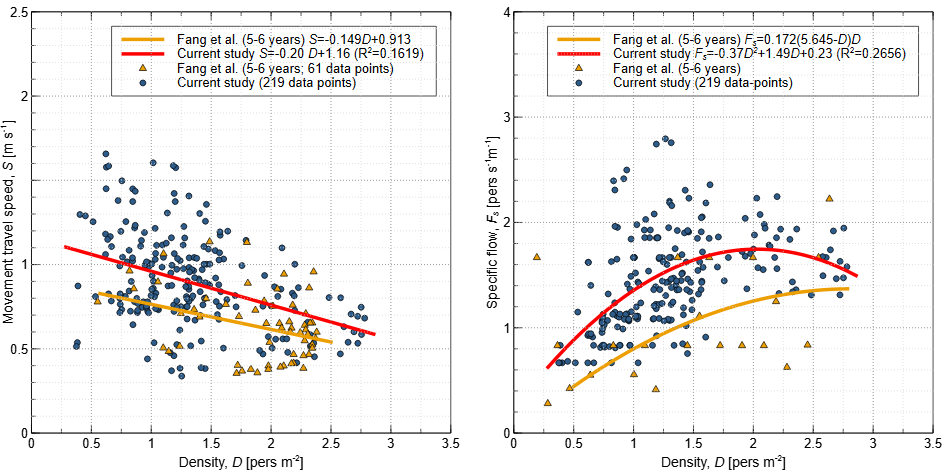}
        \vspace{-8mm}
        \caption{Movement characteristics of preschool children observed on entire straight staircases (flights and landings combined) compared to Fang et al.~\cite{fang_experimental_2019}. \textbf{Left}: Travel speeds;   
        \textbf{Right}: Specific flows.}
    \label{fig:speed_stair_compar}
\end{figure}

\begin{figure}
    \centering
        \includegraphics[width=1.0\textwidth]{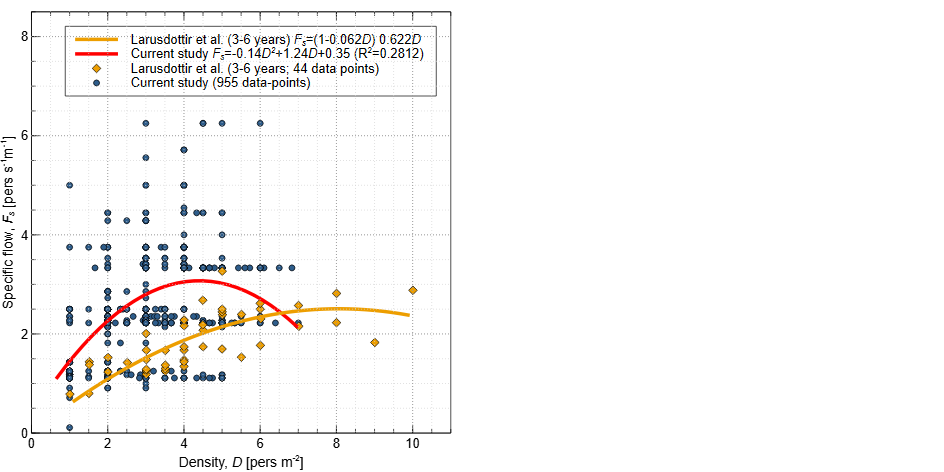}
        \vspace{-8mm}
        \caption{Specific flows per density [pers$\cdot$m$^{-2}$] in doorways compared to Larusdottir and Dederichs \cite{larusdottir_evacuation_2014}.}
    \label{fig:flow_door_compar}
\end{figure}

The presented speed- and flow-density data sets for all observations were compared with the experimental data describing evacuation of preschool children available in the literature \cite{fang_experimental_2019,larusdottir_evacuation_2014}. For comparisons, quantitatively described trend lines were adopted from the literature and graphically supported by scattered data points digitised from the original graphs \cite{fang_experimental_2019, larusdottir_evacuation_2014}.  
Figure~\ref{fig:speed_cor_comparII} provides a comparison of the presented speed-density and flow-density relationships observed for children walking in corridors with existing research reported by \cite{fang_experimental_2019}. Although the observations in the corridors in this study are limited to low density conditions (that is, approx. 1.9~children$\cdot$m$^{-2}$), the trends indicated for speed-density data in both studies are in general agreement (see Figure\ref{fig:speed_cor_comparII} Left). As shown in Figure\ref{fig:speed_cor_comparII} (Right), higher specific flows (to 4~pers$\cdot$s$^{-1}$\,m$^{-1}$) under low density conditions were observed in corridors in this study. Therefore, despite similar peak values between 2.0 and 2.5~pers$\cdot$s$^{-1}$\,m$^{-1}$, the application ranges of the compared trend lines differ considerably. 
Fang et al.~\cite{fang_experimental_2019} also provided experimental data that described movement of children on entire straight staircases with two flights and one landing (see Figure~\ref{fig:speed_stair_compar} for comparison with the current data sets). In this case, the density intervals for the plotted data sets are comparable (to approximately 3~children$\cdot$m$^{-2}$) and the provided trends appear to show similar shapes and slopes. Separate results for flights and landings on straight staircases and for spiral staircases were not identified in our literature review. 
Figure~\ref{fig:flow_door_compar} shows a graphical comparison of the flow-density relationships describing the movement of preschool children through the doors reported in \cite{larusdottir_evacuation_2014} with the results presented here. The specific flows observed in the evacuation drills presented reach higher values (the maximum value of the fitted curve is approximately 3~pers$\cdot$s$^{-1}$\,m$^{-1}$) and are limited to lower density intervals (to approximately 7~children$\cdot$m$^{-2}$). \par 
Several findings describing children's behaviour during the movement phase of the evacuation process (mostly focused on staircase movement) can be found in previous research studies \cite{capote_children_2012, larusdottir_evacuation_2012, larusdottir_evacuation_2014, najmanova_experimental_2017}. A summary of results from this study in relation to those findings is summarised in Table~\ref{tab:behaviour_mov}. 
\begin{table}[hbt!]
    \centering
    \footnotesize
    \begin{tabular}{R{9cm}|R{5cm}}
    \hline
    \textbf{Finding [references]}   &  \textbf{The current study } \\
      \hline
     Handrails are frequently used by preschool children when climbing downstairs \cite{capote_children_2012, larusdottir_evacuation_2014, najmanova_experimental_2017} & Confirmed \\
      \hline
     Preschool children's movement on staircases is adversely affected by a marking time pattern  \cite{capote_children_2012, najmanova_experimental_2017} & Confirmed \\
     \hline
    Children's behaviour on staircases may be affected by their familiarity with the escape route, staircase geometry, and environment \cite{larusdottir_evacuation_2012, larusdottir_evacuation_2014, najmanova_experimental_2017} & Partially confirmed \\
    \hline
    \end{tabular}
    \caption{Summary of behavioural aspects considering the movement behaviour of preschool children presented in this study and in the literature.}
    \label{tab:behaviour_mov}
    \end{table}
Larusdottir et al.~\cite{larusdottir_evacuation_2014} reported that 58.6\% of children three to six years of age used handrails on staircases for the evacuation drills they observed. Capote et al.~\cite{capote_children_2012} mentioned that 72\% of children from four to six years of age (66\% of children in evacuation drill that followed) used handrails in an evacuation drill. The results presented here revealed that 67.1\% of children used handrails (of this standard handrails, 40.5\%; children's handrails, 22.3\%; balusters, 4.2\%) when going downstairs and these findings confirmed the hypothesis that handrails on staircases represent important and frequently used aid components. \par
Capote et al.~\cite{capote_children_2012} concluded that 55.8\% of children 4--6~years of age (33.3\% in evacuation drill that followed) used a marking time pattern when walking downstairs on straight staircase. Comparing these values to results from this study, only 9.8\% of all children (19.8\% of Junior children) were observed using a marking time pattern when walking downstairs on straight staircases. However, our observations confirmed that children who used a marking time pattern were less confident, slower, and required more frequent staff member/older children's assistance when going downstairs. \par %
Although this study did not examine the impact of different environments or the familiarity of partcipants with escape routes, the influence of staircase geometry on the children's movement behaviour was confirmed for spiral staircases, where two travel paths with considerably different slopes and sizes were used by children walking in pairs. which would not be expected by adults. Children going downstairs on the steeper and narrower travel parts were slower, stepped more carefully, and used handrails more often. \par
Larusdottir et al.~\cite{larusdottir_evacuation_2014} reported on assistance required by preschool children on staircases. They used \quotes{No physical assistance}, \quotes{Some physical assistance} (including hand holding), and \quotes{Carried} categories. Their results for children three to six years of age are summarised in Table~\ref{tab:assistance_stair} and compared with this study's results obtained for all age groups (straight and spiral staircases included). Less physical assistance was provided to children during the evacuation drills in this study, perhaps due to local evacuation procedures and staff members strategies. Observations revealed that the children went downstairs mostly independently, following verbal instructions from staff members who were helping the slowest and least experienced children. Almost no physical support was provided to children on spiral staircases due to the limited space conditions. \par
\begin{table}[hbt!]
    \centering
    \footnotesize
    \begin{tabular}{R{4.0cm}|R{2.5cm}|R{4.0cm}|C{2.5cm}}
    \hline
     \multicolumn{2}{C{6.5cm}|}{\textbf{Larusdottir et al.~\cite{larusdottir_evacuation_2014}}}   & \multicolumn{2}{C{6.5cm}}{\textbf{Current study}} \\
     \hline 
      Assistance & Frequency [\%] & Assistance & Frequency [\%]  \\
      \hline
      No physical assistance & 85.2 & No physical assistance & 93.9 \\
      \hline
      \multirow{2}{=}{Some physical assistance} & \multirow{2}{=}{13.9} & Gentle pushing  & 1.3 \\
      \cline{3-4}
      & & Physical contact needed and hand holding & 5.5 \\
      \hline
      Carried & 0.9 & Carried & 0.0 \\
      \hline
      \end{tabular}
\caption{Comparison of levels of assistance required by children during the movement phase (on staircases) presented in this study and the literature.}
    \label{tab:assistance_stair}
\end{table}
\subsection{Behavioural statements}\label{sec:statements}
Specifics of evacuation procedures and the behaviour of the children observed during the movement phase in this study are presented using the concept of behavioural statements \cite{gwynne_guidance_2016,kuligowski_guidance_2017}. To enable a deeper understanding and quantification of children's movement in engineering applications, a set of nine behavioural statements about the movement evacuation behaviour of preschool children in nursery schools were formulated and provided in Table~\ref{tab:statement}. Detailed interpretation of the behavioural statements considering physical movement characteristics and behavioural itineraries is presented in the following subsections. Further implementations of the behavioural statements in evacuation design can be supported by the relevant data presented in the related parts of this paper.
\begin{table}[hbt!]
    \centering
    \footnotesize
    \begin{tabular}{R{0.4cm}|R{14cm}}
    \hline
    \textbf{\#} & \textbf{behavioural statement} \\
    \hline
    1 & Movement abilities of children differ from those of adults and are age-dependent. \\
    \hline
    2 & Going down staircases is the most challenging task for preschool children travelling through buildings. Junior children under 4~years of age may move differently than older children because of their limited motor abilities, often forcing them to use handrails or exhibit marking time pattern.  \\
    \hline
    3 & Preschool children use handrails frequently when they move on staircases. The presence or absence of handrails may considerably influence children's selection of travel paths on staircases. \\
    \hline
    4 & Close physical contact (such as touching, pushing, hand holding) with each other is typical for preschool children when travelling through buildings. Under higher density conditions, children can form very compact crowds with minimal body buffer zones. \\
    \hline
    5 & Compared to adults, preschool children can perceive movement as a kind of game resulting in use of various movement patterns such as racing, jumping, and hopping. \\
    \hline
    6 & When unexpected barriers occurred on evacuation routes (e.g.,~an opened door reducing the effective width of a route, chairs located in the middle of a walkway), preschool children prefer squeezing through the narrow space and form bottlenecks to navigate the obstacles. \\
    \hline
    7 & Preschool children travelling through buildings without constant staff member supervision tend to use familiar escape routes. \\
    \hline
    8 & Before leaving a building, preschool children appear to follow daily routines and practices (e.g.,~putting on shoes and clothing). \\
    \hline
    9 & The movement behaviour of preschool children, including hand holding, moving in pairs, moving in compacted groups, using handrails, using doors independently (i.e.,~opening, unlocking them), can be strongly influenced by staff member instructions as well as daily routines, rules, and practises employed in each nursery school. \\    
     \hline
    \end{tabular}
\caption{Overview of the behavioural statements for the movement evacuation behaviour of preschool children in nursery schools.}
    \label{tab:statement}
\end{table}
\subsubsection{Physical movement characteristics}
The findings from this study suggest that, in engineering applications, different movement characteristics should be assumed for preschool children at different ages. Table~\ref{tab:mov_eng} presents movement characteristics observed in this study for children at low density conditions (0.00--0.05~m$^2$\,m$^{-2}$) related to two simplified age groups: \quotes{Junior under 4~years} and \quotes{Senior over 4~years} (the latter group includes the results for both Senior and Senior+ classes; Mixed classes were not included). 

\begin{table}[hbt!]
    \centering
    \footnotesize
    \begin{tabular}{R{5cm}|C{4.5cm}|R{4.5cm}}
    \hline
    \multirow{2}{=}{\textbf{Evacuation route}} &  \multicolumn{2}{R{9cm}}{\textbf{Travel speed \newline (mean/min/max/SD) [m$\cdot$s$^{-1}$] (data points)}} \\
    \cline{2-3}
    & \textbf{Junior under 4~years} & \textbf{Senior over 4~years} \\
    \hline
    Corridors (W) &  1.06 / 0.42 / 1.85 / 0.26 (139) & 1.43 / 0.54 / 3.00 / 0.59 (147) \\
    \hline
    Corridors (R) & 1.82 / 1.12 / 2.78 / 0.42 (47) & 2.98 / 1.03 / 5.25 / 0.84 (177) \\
    \hline
    Straight staircases: flights & 0.71 / 0.28 / 1.21 / 0.21 (132) & 0.94 / 0.33 / 1.74 / 0.26 (412) \\
    \hline
    Straight staircases: landings (W) & 0.84 / 0.17 / 1.31 / 0.28 (43) & 1.03 / 0.36 / 1.86 / 0.40 (43) \\
    \hline
    Straight staircases: landings (R) & 1.65 / 1.45 / 1.86 / 0.17 (6) & 1.47 / 0.44 / 2.55 / 0.56 (39) \\
    \hline
    Straight staircases: whole & 0.77 / 0.34 / 1.11 / 0.17 (49) &  1.03 / 0.37 / 1.66 / 0.27 (108) \\
    \hline
    Doorways (W) & 0.92 / 0.40 / 1.59 / 0.12 (35) & 1.09 / 0.67 / 1.35 / 0.19 (25) \\
    \hline
    Doorways (R) & 1.70 / 1.03 / 2.78 / 0.48 (13) & 1.64 / 0.83 / 3.45 / 0.52 (29) \\
    \hline
        \end{tabular}
    \caption{Travel speeds measured under low density conditions (0.00--0.05~m$^2$\,m$^{-2}$) for children in two age subgroups (Mixed classes excluded); W: walking travel speeds, R: running travel speeds}
    \label{tab:mov_eng}
\end{table}
To a large extent, the movement characteristics of preschool children can be defined by various behavioural aspects presented here as behavioural statements (\# 1--9 in Table~\ref{tab:statement}). The following factors influencing evacuation processes should be properly considered when selecting specific values for movement characteristics of children in engineering applications: children's motivation to move, level of supervision, instructions given by responsible staff members, daily routines known by children, and educational practices. The travel speeds for children moving in pairs and holding each other's hands can be impeded, for example, by a slower child in a pair who does not have a handrail on one side of a staircase. \par 
The movement characteristics of preschool children also differ from those of the older children and adults due to their different physical dimensions and the personal space they require. In Table~\ref{tab:dimension_eng}, a summary of the main characteristics of preschool children including shoulder width, chest depth, and occupied area is provided for the two suggested age subgroups together with the reference values commonly accepted for adults \cite{predtechenskii_planning_1978}. This comparison confirms the importance of using relevant density units (m$^2$\,m$^{-2}$) when comparing the movement of children and adults, because density expressions in units of pers$\cdot$m$^{-2}$ may lead to misleading interpretations.

\begin{table}[hbt!]
    \centering
    \footnotesize
    \begin{tabular}{R{3.2cm}|R{4cm}|R{4cm}|R{1.5cm}}
    \hline
    \textbf{Characteristic} & \textbf{Junior under 4~years} \newline range of means (average value) & \textbf{Senior over 4~years} \newline range  of means (average value) & \textbf{Adults} \cite{predtechenskii_planning_1978}\\
    \hline
    Shoulder width [cm] & 22.0--22.8 (22.4) & 23.8--25.1 (24.5) & 48 \\
    \hline
    Chest depth [cm] &  13.7--13.9 (13.8)  & 14.3--14.7 (14.5) & 30 \\
    \hline
    Occupied area [m$^2$] & 0.024--0.025 (0.025) & 0.027--0.029 (0.028) & 0.113 \\
    \hline
    \end{tabular}
\caption{Physical dimensions and assumed occupied area for preschool children in two age subgroups according to \cite{kobzova_6th_2004} compared to those generally accepted for adults \cite{predtechenskii_planning_1978}.}
    \label{tab:dimension_eng}
\end{table}
\subsubsection{Behavioural itineraries}
Several challenges related to the specifics of the evacuation movement of preschool children can emerge in evacuation predictions and modelling. Observations from this study revealed that the children frequently moved in organized, compact, and supervised groups, whose shapes (e.g., single file, a group with pairs, a huddle group) varied according to daily routines and actual instructions given by staff members. If children moved independently on evacuation routes without supervision, they could easily go in an unexpected direction or miss an exit. The children often formed pairs holding each other's hands, and as such, overtaking behaviours were not often possible, and individual travel speeds were limited not only by density conditions but also by the movement of the children in front of them. In the observed evacuation drills, it was not uncommon for a group of children to split into subgroups or wait for slower children. Due to a higher level of organization in some nursery schools, where children were instructed to stay together in a compact group and not mix with other classes, this typically resulted in stopping and yielding the way to other groups (notably on staircases and in doorways). Additional stops were required to open or unlock doors or due to waiting for staff members delayed as they assisted less experienced children. Since most of these factors are difficult for fire safety engineers to foresee, it is challenging to predict the evacuation movement of preschool children in nursery schools. Therefore, it can be recommended that the impact of all the behavioural statements presented is considered together and the analysis of uncertainties related to the variability of evacuation and movement scenarios is supported by the means of sensitivity analysis and/or probabilistic methods. \par
\section{Conclusions}
Preschool children represent a vulnerable segment of the population whose potentially limited self-rescue capabilities can result in specific requirements during emergency evacuations. This paper expands our current understanding of the movement and behaviour of preschool children (three to seven years of age) by analysing 15~evacuation drills conducted in ten participating nursery schools in the Czech Republic. Our observations showed that the movement characteristics of preschool children were age dependent, and children's different ages and abilities should be properly considered in fire safety design. This study provides valuable data which can be used to calibrate existing evacuation models and to develop evacuation models towards a more accurate representation of preschool children movement and behaviour. \par

\section*{Acknowledgments}
The authors thank all participants involved in the evacuation drills for their efforts and cooperation, which made this research project possible. English language editorial guidance was provided by Dr. Stephanie Krueger. The authors also thank Dr. Petr Novák for constructive criticism of the manuscript.

\section*{CRediT authorship contribution statement}
\textbf{Hana Najmanov\'a}: Conceptualization, \sep 
Methodology, \sep 
Formal analysis, \sep 
Investigatio,n \sep 
Resources, \sep 
Data Curation, \sep 
Writing - Original Draft; \sep 
\textbf{Enrico Ronchi}: Supervision, \sep 
Methodology, \sep 
Writing - Review \& Editing


\bibliography{Hana_zotero.bib}

\appendix
\clearpage\newpage
\section{Complementary material for the results on movement in corridors}
\label{app:cor}
\setcounter{table}{0}
\renewcommand\thetable{\Alph{section}.\arabic{table}}
\begin{table}[hbt!]
    \centering
    \footnotesize
    \begin{tabular}{R{1.3cm}|R{1.5cm}|R{5.3cm}|R{5.3cm}}
    \hline
      \textbf{Age group} & \textbf{Density interval [m$^2$\, m$^{-2}$]}  & \textbf{Travel speed \newline (mean/min/max/SD) [m$\cdot$s$^{-1}$] (data points)} & \textbf{Density \newline (mean/min/max/SD) [m$^2$\,m$^{-2}$] (data points)} \\
    \hline
    Junior & 0.00--0.05 & 1.25 / 0.42 / 2.78 / 0.45 (186) & 0.02 / 0.01 / 0.05 / 0.01 (186)  \\
    \hline
    \multirow{2}{=}{Senior} & 0.00--0.05 &  1.97 / 0.54 / 4.36 / 0.92 (236) &  0.02 / 0.01 / 0.05 / 0.01 (236)  \\
    \cline{2-4}
    & 0.06--0.10 & 0.88 / 0.66 / 1.08 / 0.12 (21) & 0.08 / 0.07 / 0.09 / 0.01 (21)  \\
    \hline
    Senior+ & 0.00--0.05 & 3.10 / 1.03 / 5.25 / 1.01 (88)  & 0.01 / 0.00 / 0.03 / 0.01 (88) \\
    \hline
    Mixed  & 0.00--0.05 & 2.02 / 0.39 / 4.65 / 0.74 (176) & 0.02 / 0.00 / 0.05 / 0.01 (176) \\
    \hline
    \multicolumn{2}{l|}{\textbf{Total}}  & \multicolumn{1}{l|}{\textbf{1.90 / 0.39 / 5.25 / 0.96 (707)}} & \multicolumn{1}{l}{\textbf{0.02 / 0.00 / 0.09 / 0.01 (707)}} \\
    \hline
    \end{tabular}
    \caption{Travel speeds for children measured in the corridors during the evacuation drills (the values in the row \quotes{Total} indicated the results over the total sample, i.e.,~all age groups and density intervals).}
    \label{tab:speed_cor}
\end{table}
\begin{table}[hbt!]
    \centering
    \footnotesize
    \begin{tabular}{R{1.3cm}|R{1.5cm}|R{5.3cm}|R{5.3cm}}
    \hline
      \textbf{Age group} & \textbf{Density interval [m$^2$\,m$^{-2}$]}  & \textbf{Travel speed \newline (mean/min/max/SD) [m$\cdot$s$^{-1}$] (data points)} & \textbf{Density \newline (mean/min/max/SD) [m$^2$\,m$^{-2}$] (data points)} \\
    \hline
    Junior & 0.00--0.05 & 1.06 / 0.42 / 1.85 / 0.26 (139) &  0.03 / 0.01 / 0.05 / 0.01 (139) \\
    \hline
    \multirow{2}{=}{Senior} & 0.00--0.05 & 1.38 / 0.54 / 3.00 / 0.57 (132) & 0.03 / 0.01 / 0.05 / 0.01 (132)  \\
    \cline{2-4}
        & 0.06--0.10 & 0.88 / 0.66 / 1.08 / 0.12 (21) & 0.08 / 0.07 / 0.09 / 0.01 (21)  \\
    \hline
    Senior+ & 0.00--0.05 & 1.92 / 1.40 / 2.98 / 0.55 (15) & 0.02 / 0.01 / 0.03 / 0.01 (15)  \\
    \hline
    Mixed & 0.00--0.05 & 1.55 / 0.39 / 3.22 / 0.68 (50) & 0.02 / 0.01 / 0.04 / 0.01 (50) \\
    \hline
    \multicolumn{2}{l|}{\textbf{Total}}  & \multicolumn{1}{l|}{\textbf{1.27 / 0.39 / 3.22 / 0.53 (357)}} & \multicolumn{1}{l}{\textbf{0.03 / 0.01 / 0.09 / 0.02 (357)}} \\
    \hline
    \end{tabular}
    \caption{Walking travel speeds for children measured in the corridors during the evacuation drills (the values in the row \quotes{Total} indicated the results over the total sample, i.e.,~all age groups and density intervals).}
    \label{tab:speed_cor_walk}
\end{table}
\begin{table}[hbt!]
    \centering
    \footnotesize
    \begin{tabular}{R{1.3cm}|R{1.5cm}|R{5.3cm}|R{5.3cm}}
    \hline
      \textbf{Age group} & \textbf{Density interval [m$^2$\,m$^{-2}$]}  & \textbf{Travel speed \newline (mean/min/max/SD) [m$\cdot$s$^{-1}$] (data points)} & \textbf{Density \newline (mean/min/max/SD) [m$^2$\,m$^{-2}$] (data points)} \\
    \hline
    Junior & 0.00--0.05 & 1.82 / 1.12 / 2.78 / 0.42 (47) & 0.02 / 0.01 / 0.05 / 0.01 (47) \\
    \hline
    Senior & 0.00--0.05 & 2.73 / 1.27 / 4.36 / 0.69 (104) &  0.02 / 0.01 / 0.04 / 0.01 (104)  \\
    \hline
    Senior+ & 0.00--0.05 & 3.35 / 1.03 / 5.25 / 0.91 (73) & 0.01 / 0.00 / 0.03 / 0.01 (73)  \\
    \hline
    Mixed & 0.00--0.05 & 2.21 / 1.11 / 4.65 / 0.68 (126) & 0.02 / 0.00 / 0.05 / 0.01 (126) \\
    \hline
    \multicolumn{2}{l|}{\textbf{Total}}  & \multicolumn{1}{l|}{\textbf{2.55 / 1.03 / 5.25 / 0.87 (350)}} & \multicolumn{1}{l}{\textbf{0.02 / 0.00 / 0.05 / 0.01 (350)}} \\
    \hline
    \end{tabular}
    \caption{Running travel speeds for children measured in corridors during the evacuation drills (the values in the row \quotes{Total} indicated the results over the total sample, i.e.,~all age groups and density intervals).}
    \label{tab:speed_cor_run}
\end{table}
\begin{table}[hbt!]
    \centering
    \footnotesize
    \begin{tabular}{R{1.3cm}|R{1.5cm}|R{5.3cm}|R{5.3cm}}
    \hline
      \textbf{Age group}  & {\textbf{Density interval [m$^2$\,m$^{-2}$]}} & \textbf{Specific flow \newline (mean/min/max/SD) [pers$\cdot$s$^{-1}$\,m$^{-1}$] (data points)} & \textbf{Density (mean/min/max/SD) [m$^2$\,m$^{-2}$] (data points)} \\
    \hline
    Junior & 0.00--0.05 & 1.72 / 0.65 / 4.17 / 0.80 (186) & 0.02 / 0.01 / 0.05 / 0.01 (186)  \\
    \hline
    \multirow{2}{=}{Senior}  & 0.00--0.05 & 2.00 / 0.67 / 4.17 / 0.90 (236) &  0.02 / 0.01 / 0.05 / 0.01 (236)  \\
     \cline{2-4}
       & 0.06--0.10 & 1.68 / 1.33 / 2.67 / 0.54 (21) & 0.08 / 0.07 / 0.09 / 0.01 (21)  \\
    \hline
    Senior+ & 0.00--0.05 & 1.61 / 0.83 / 3.33 / 0.79 (88)  &  0.01 / 0.00 / 0.03 / 0.01 (88)  \\
    \hline
    Mixed & 0.00--0.05 & 1.63 / 0.67 / 4.17 / 0.79 (176) & 0.02 / 0.00 / 0.05 / 0.01 (176) \\
    \hline
    \multicolumn{2}{l}{\textbf{Total}} & \multicolumn{1}{l}{\textbf{1.78 / 0.65 / 4.17 / 0.84 (707)}} & \multicolumn{1}{l}{\textbf{0.02 / 0.00 / 0.09 / 0.01 (707)}} \\
    \hline
    \end{tabular}
    \caption{Specific flows for children measured in the corridors during the evacuation drills (the values in the row \quotes{Total} indicated the results over the total sample, i.e.,~all age groups and density intervals).}
    \label{tab:flow_cor}
\end{table}

\clearpage\newpage
\section{Complementary material for the results on movement on straight staircases}
\label{app:stair}
\setcounter{table}{0}
\renewcommand\thetable{\Alph{section}.\arabic{table}}
\begin{table}[hbt!]
    \centering
    \footnotesize
    \begin{tabular}{R{1.3cm}|R{1.5cm}|R{5.3cm}|R{5.3cm}}
    \hline
      \textbf{Age group} & \textbf{Density interval [m$^2$\,m$^{-2}$]} & \textbf{Travel speed \newline (mean/min/max/SD) [m$\cdot$s$^{-1}$] (data points)} & \textbf{Density (mean/min/max/SD) [m$^2$\,m$^{-2}$] (data points)} \\
    \hline
    \multirow{2}{=}{Junior} & 0.00--0.05 & 0.71 / 0.28 / 1.21 / 0.21 (132) & 0.03 / 0.01 / 0.05 / 0.01 (132)  \\
    \cline{2-4}
       & 0.06--0.10 & 0.77 / 0.49 / 1.13 / 0.17 (27) & 0.06 / 0.05 / 0.06 / 0.01 (27)  \\
    \hline
    \multirow{2}{=}{Senior} & 0.00--0.05 & 0.94 / 0.33 / 1.74 / 0.27 (346) &  0.03 / 0.01 / 0.05 / 0.01 (346)  \\
    \cline{2-4}
       & 0.06--0.10 & 0.78 / 0.44 / 1.17 / 0.19 (45) & 0.06 / 0.05 / 0.07 / 0.01 (45)  \\
    \hline
    Senior+ & 0.00--0.05 & 0.94 / 0.65 / 1.35 / 0.18 (66)  & 0.02 / 0.01 / 0.04 / 0.01 (66) \\
    \hline
    \multirow{2}{=}{Mixed} & 0.00--0.05 & 0.73 / 0.32 / 1.62 / 0.26 (128) & 0.02 / 0.01 / 0.05 / 0.01 (128) \\
    \cline{2-4}
       & 0.06--0.10 & 0.79 / 0.53 / 1.14 / 0.30 (4) & 0.05 / 0.05 / 0.05 / 0.00 (4)  \\
    \hline
    \multicolumn{2}{l}{\textbf{Total}} & \multicolumn{1}{l}{\textbf{0.85 / 0.28 / 1.74 / 0.26 (748)}} & \multicolumn{1}{l}{\textbf{0.03 / 0.01 / 0.07 / 0.01 (748)}} \\
    \hline
    \end{tabular}
    \caption{Travel speeds for children measured on the flights of the straight staircases during the evacuation drills (the values in the row \quotes{Total} indicated the results over the total sample, i.e.,~all age groups and density intervals).}
    \label{tab:speed_stair_fl}
\end{table}
\begin{table}[hbt!]
    \centering
    \footnotesize
    \begin{tabular}{R{1.3cm}|R{1.5cm}|R{5.3cm}|R{5.3cm}}
    \hline
      \textbf{Age group}  & {\textbf{Density interval [m$^2$\,m$^{-2}$]}} & \textbf{Specific flow \newline (mean/min/max/SD) [pers$\cdot$s$^{-1}$\,m$^{-1}$] (data points)} & \textbf{Density (mean/min/max/SD) [m$^2$\,m$^{-2}$] (data points)} \\
    \hline
    Junior & 0.00--0.05 & 1.13 / 0.45 / 3.57 / 0.53 (132) & 0.03 / 0.01 / 0.05 / 0.01 (132)  \\
        \cline{2-4}
    & 0.06--0.10 & 1.64 / 0.71 / 3.57 / 0.88 (27) & 0.06 / 0.05 / 0.06 / 0.01 (27)  \\
    \hline
    \multirow{2}{=}{Senior}  & 0.00--0.05 & 1.49 / 0.50 / 2.73 / 0.58 (346) &  0.03 / 0.01 / 0.05 / 0.01 (346)  \\
     \cline{2-4}
       & 0.06--0.10 & 1.64 / 0.83 / 2.50 / 0.55 (45) & 0.06 / 0.05 / 0.07 / 0.01 (45)  \\
    \hline
    Senior+ & 0.00--0.05 & 0.72 / 0.53 / 1.58 / 0.29 (66)  &  0.02 / 0.01 / 0.04 / 0.01 (66)  \\
    \hline
    Mixed & 0.00--0.05 & 0.91 / 0.53 / 2.00 / 0.43 (128) & 0.02 / 0.01 / 0.05 / 0.01 (128) \\
        \cline{2-4}
    & 0.06--0.10 & 1.81 / 1.25 / 2.00 / 0.38 (4) & 0.05 / 0.05 / 0.06 / 0.01 (4)  \\
    \hline
    \multicolumn{2}{l}{\textbf{Total}} & \multicolumn{1}{l}{\textbf{1.28 / 0.45 / 3.57 / 0.61 (748)}} & \multicolumn{1}{l}{\textbf{0.03 / 0.01 / 0.07 / 0.01 (748)}} \\
    \hline
    \end{tabular}
    \caption{Specific flows for children measured on the flights of the straight staircases during the evacuation drills (the values in the row \quotes{Total} indicated the results over the total sample, i.e.,~all age groups and density intervals).}
    \label{tab:flow_stair_fl}
\end{table}
\begin{table}[hbt!]
    \centering
    \footnotesize
    \begin{tabular}{R{1.3cm}|R{1.5cm}|R{5.3cm}|R{5.3cm}}
    \hline
      \textbf{Age group} & \textbf{Density interval [m$^2$\,m$^{-2}$]} & \textbf{Travel speed \newline (mean/min/max/SD) [m$\cdot$s$^{-1}$] (data points)} & \textbf{Density (mean/min/max/SD) [m$^2$\,m$^{-2}$] (data points)} \\
    \hline
    Junior & 0.00--0.05 & 0.94 / 0.17 / 1.86 / 0.38 (49) & 0.02 / 0.00 / 0.04 / 0.01 (49)  \\
    \hline
    \multirow{3}{=}{Senior} & 0.00--0.05 & 1.24 / 0.36 / 2.55 / 0.53 (82) &  0.03 / 0.01 / 0.05 / 0.01 (82)  \\
    \cline{2-4}
       & 0.06--0.10 & 0.73 / 0.22 / 1.60 / 0.34 (73) & 0.07 / 0.05 / 0.10 / 0.01 (73)  \\
           \cline{2-4}
       & 0.11--0.15 & 0.51 / 0.22 / 0.83 / 0.19 (29) & 0.12 / 0.10 / 0.14 / 0.01 (29)  \\
    \hline
    \multirow{2}{=}{Mixed} & 0.00--0.05 & 1.31 / 0.56 / 2.25 / 0.52 (19) & 0.02 / 0.01 / 0.05 / 0.01 (19) \\
     \cline{2-4}
       & 0.06--0.10 & 1.20 / 0.75 / 1.54 / 0.34 (4) & 0.05 / 0.05 / 0.06 / 0.01 (4)  \\
    \hline
    \multicolumn{2}{l}{\textbf{Total}} & \multicolumn{1}{l}{\textbf{0.96 / 0.17 / 2.55 / 0.50 (256)}} & \multicolumn{1}{l}{\textbf{0.05 / 0.00 / 0.14 / 0.03 (256)}} \\
    \hline
    \end{tabular}
    \caption{Travel speeds for children measured on the landings of the straight staircases during the evacuation drills (the values in the row \quotes{Total} indicated the results over the total sample, i.e.,~all age groups and density intervals).}
    \label{tab:speed_stair_land}
\end{table}
\begin{table}[hbt!]
    \centering
    \footnotesize
    \begin{tabular}{R{1.3cm}|R{1.5cm}|R{5.3cm}|R{5.3cm}}
    \hline
      \textbf{Age group} & \textbf{Density interval [m$^2$\,m$^{-2}$]} & \textbf{Travel speed \newline (mean/min/max/SD) [m$\cdot$s$^{-1}$] (data points)} & \textbf{Density (mean/min/max/SD) [m$^2$\,m$^{-2}$] (data points)} \\
    \hline
    Junior & 0.00--0.05 & 0.84 / 0.17 / 1.31 / 0.28 (43) &  0.02 / 0.00 / 0.04 / 0.01 (43) \\
    \hline
    \multirow{3}{=}{Senior} & 0.00--0.05 & 1.03 / 0.36 / 1.86 / 0.40 (43) & 0.03 / 0.01 / 0.05 / 0.01 (43)  \\
    \cline{2-4}
       & 0.06--0.10 & 0.68 / 0.22 / 1.60 / 0.34 (56) & 0.07 / 0.05 / 0.10 / 0.02 (56)  \\
    \cline{2-4}
       & 0.11--0.15 & 0.51 / 0.22 / 0.83 / 0.19 (29) & 0.12 / 0.10 / 0.14 / 0.01 (29)  \\
    \hline
    Mixed & 0.00--0.05 & 0.65 / 0.59 / 0.70 / 0.05 (3) & 0.02 / 0.02 / 0.03 / 0.01 (3) \\
    \hline
    \multicolumn{2}{l}{\textbf{Total}} & \multicolumn{1}{l}{\textbf{0.78 / 0.17 / 1.86 / 0.37 (174)}} & \multicolumn{1}{l}{\textbf{0.06 / 0.00 / 0.14 / 0.04 (174)}} \\
    \hline
    \end{tabular}
    \caption{Walking travel speeds for children measured on the landings of straight staircases during the evacuation drills (the values in the row \quotes{Total} indicated the results over the total sample, i.e.,~all age groups and density intervals).}
    \label{tab:speed_stair_str_walk}
\end{table}
\begin{table}[hbt!]
    \centering
    \footnotesize
    \begin{tabular}{R{1.3cm}|R{1.5cm}|R{5.3cm}|R{5.3cm}}
    \hline
      \textbf{Age group} & \textbf{Density interval [m$^2$\,m$^{-2}$]} & \textbf{Travel speed \newline (mean/min/max/SD) [m$\cdot$s$^{-1}$] (data points)} & \textbf{Density (mean/min/max/SD) [m$^2$\,m$^{-2}$] (data points)} \\
    \hline
    Junior & 0.00--0.05 & 1.65 / 1.45 / 1.86 / 0.17 (6) & 0.02 / 0.01 / 0.03 / 0.00 (6) \\
    \hline
    \multirow{2}{=}{Senior} & 0.00--0.05 & 1.47 / 0.44 / 2.55 / 0.56 (39) &  0.03 / 0.01 / 0.05 / 0.01 (39)  \\
        \cline{2-4}  
       & 0.06--0.10 & 0.89 / 0.43 / 1.43 / 0.31 (17) & 0.06 / 0.05 / 0.08 / 0.01 (17)  \\
    \hline
    \multirow{2}{=}{Mixed} & 0.00--0.05 & 1.44 / 0.56 / 2.25 / 0.47 (16) & 0.02 / 0.01 / 0.05 / 0.01 (16) \\
     \cline{2-4}
       & 0.06--0.10 & 1.20 / 0.75 / 1.54 / 0.34 (4) & 0.05 / 0.05 / 0.06 / 0.01 (4)  \\
    \hline
    \multicolumn{2}{l}{\textbf{Total}} & \multicolumn{1}{l}{\textbf{1.35 / 0.43 / 2.55 / 0.52 (82)}} & \multicolumn{1}{l}{\textbf{0.04 / 0.01 / 0.08 / 0.02 (82)}} \\
    \hline
    \end{tabular}
    \caption{Running travel speeds for children measured on the landings of straight staircases during the evacuation drills (the values in the row \quotes{Total} indicated the results over the total sample, i.e.,~all age groups and density intervals).}
    \label{tab:speed_stair_str_run}
\end{table}
\begin{table}[hbt!]
    \centering
    \footnotesize
    \begin{tabular}{R{1.3cm}|R{1.5cm}|R{5.3cm}|R{5.3cm}}
    \hline
      \textbf{Age group}  & {\textbf{Density interval [m$^2$\,m$^{-2}$]}} & \textbf{Specific flow \newline (mean/min/max/SD) [pers$\cdot$s$^{-1}$\,m$^{-1}$] (data points)} & \textbf{Density (mean/min/max/SD) [m$^2$\,m$^{-2}$] (data points)} \\
    \hline
    Junior & 0.00--0.05 & 1.39 / 0.83 / 2.50 / 0.55 (49) & 0.02 / 0.00 / 0.04 / 0.01 (49)  \\
    \hline
    \multirow{2}{=}{Senior}  & 0.00--0.05 & 1.45 / 0.63 / 3.64 / 0.78 (82) &  0.03 / 0.01 / 0.05 / 0.01 (82)  \\
     \cline{2-4}
       & 0.06--0.10 & 1.67 / 0.64 / 3.21 / 0.50 (73) & 0.07 / 0.05 / 0.10 / 0.01 (73)  \\
     \cline{2-4}
       & 0.11--0.15 & 2.01 / 0.83 / 3.33 / 0.76 (29) & 0.12 / 0.10 / 0.14 / 0.01 (29)  \\     
    \hline
    Mixed & 0.00--0.05 & 0.77 / 0.67 / 1.33 / 0.25 (19) & 0.02 / 0.01 / 0.05 / 0.01 (19) \\
    \cline{2-4}
           & 0.06--0.10 & 1.83 / 1.33 / 2.00 / 0.33 (4) & 0.05 / 0.05 / 0.06 / 0.01 (4)  \\     
    \hline
    \multicolumn{2}{l}{\textbf{Total}} & \multicolumn{1}{l}{\textbf{1.52 / 0.63 / 3.64 / 0.69 (256)}} & \multicolumn{1}{l}{\textbf{0.05 / 0.00 / 0.14 / 0.03 (256)}} \\
    \hline
    \end{tabular}
    \caption{Specific flows for children measured on the landings of straight staircases during the evacuation drills (the values in the row \quotes{Total} indicated the results over the total sample, i.e.,~all age groups and density intervals.)}
    \label{tab:flow_stair_land}
\end{table}
\begin{table}[hbt!]
    \centering
    \footnotesize
    \begin{tabular}{R{1.3cm}|R{1.5cm}|R{5.3cm}|R{5.3cm}}
    \hline
      \textbf{Age group} & \textbf{Density interval [m$^2$\,m$^{-2}$]} & \textbf{Travel speed \newline (mean/min/max/SD) [m$\cdot$s$^{-1}$] (data points)} & \textbf{Density (mean/min/max/SD) [m$^2$\,m$^{-2}$] (data points)} \\
    \hline
    Junior & 0.00--0.05 & 0.77 / 0.34 / 1.11 / 0.17 (49) & 0.02 / 0.01 / 0.04 / 0.01 (49)  \\
    \hline
    \multirow{2}{=}{Senior} & 0.00--0.05 & 1.03 / 0.37 / 1.66 / 0.27 (108) &  0.03 / 0.01 / 0.05 / 0.01 (108)  \\
     \cline{2-4}
     & 0.06--0.10 & 0.68 / 0.42 / 1.14 / 0.18 (39) &  0.06 / 0.05 / 0.08 / 0.01 (39)  \\
    \hline
    Mixed & 0.00--0.05 & 0.90 / 0.51 / 1.30 / 0.22 (23) & 0.02 / 0.01 / 0.04 / 0.01 (23) \\
    \hline
    \multicolumn{2}{l}{\textbf{Total}} & \multicolumn{1}{l}{\textbf{0.90 / 0.34 / 1.66 / 0.27 (219)}} & \multicolumn{1}{l}{\textbf{0.04 / 0.01 / 0.08 / 0.02 (219)}} \\
    \hline
    \end{tabular}
    \caption{Travel speeds for children measured on the entire straight staircases (flights and landings combined) during the evacuation drills (the values in the row \quotes{Total} indicated the results over the total sample, i.e.,~all age groups and density intervals).}
    \label{tab:speed_stair_whole}
\end{table}
\begin{table}[hbt!]
    \centering
    \footnotesize
    \begin{tabular}{R{1.3cm}|R{1.5cm}|R{5.3cm}|R{5.3cm}}
    \hline
      \textbf{Age group}  & {\textbf{Density interval [m$^2$\,m$^{-2}$]}} & \textbf{Specific flow \newline (mean/min/max/SD) [pers$\cdot$s$^{-1}$\,m$^{-1}$] (data points)} & \textbf{Density (mean/min/max/SD) [m$^2$\,m$^{-2}$] (data points)} \\
    \hline
    Junior & 0.00--0.05 & 1.27 / 0.83 / 2.20 / 0.38 (49) & 0.02 / 0.01 / 0.04 / 0.01 (49)  \\
    \hline
    \multirow{2}{=}{Senior}  & 0.00--0.05 & 1.55 / 0.68 / 2.79 / 0.46 (108) &  0.03 / 0.01 / 0.05 / 0.01 (108)  \\
     \cline{2-4}
       & 0.06--0.10 & 1.70 / 1.31 / 2.25 / 0.25 (39) & 0.06 / 0.05 / 0.08 / 0.01 (39)  \\
     \hline
    Mixed & 0.00--0.05 & 0.98 / 0.67 / 1.75 / 0.31 (23) & 0.02 / 0.01 / 0.04 / 0.01 (23) \\
    \hline
    \multicolumn{2}{l}{\textbf{Total}} & \multicolumn{1}{l}{\textbf{1.45 / 0.67 / 2.79 / 0.45 (219)}} & \multicolumn{1}{l}{\textbf{0.04 / 0.01 / 0.08 / 0.02 (219)}} \\
    \hline
    \end{tabular}
    \caption{Specific flows ofor children measured on the entire straight staircases during the evacuation drills (the values in the row \quotes{Total} indicated the results over the total sample, i.e.,~all age groups and density intervals).}
    \label{tab:flow_stair_whole}
\end{table}
\clearpage\newpage
\section{Complementary material for the results on movement on spiral staircases}
\label{app:stair_spiral}
\setcounter{table}{0}
\renewcommand\thetable{\Alph{section}.\arabic{table}}
\begin{table}[hbt!]
    \centering
    \footnotesize
    \begin{tabular}{R{1.3cm}|R{1.5cm}|R{5.3cm}|R{5.3cm}}
    \hline
      \textbf{Age group} & \textbf{Density interval [m$^2$\,m$^{-2}$]} & \textbf{Travel speed \newline (mean/min/max/SD) [m$\cdot$s$^{-1}$] (data points)} & \textbf{Density (mean/min/max/SD) [m$^2$\,m$^{-2}$] (data points)} \\
    \hline
    \multirow{2}{=}{Mixed} & 0.00--0.05 & 0.79 / 0.31 / 1.41 / 0.25 (107) &  0.04 / 0.02 / 0.05 / 0.01 (107)  \\
     \cline{2-4}
     & 0.06--0.10 & 0.61 / 0.27 / 1.29 / 0.21 (130) &  0.07 / 0.05 / 0.10 / 0.01 (130)  \\
    \hline
    \multicolumn{2}{l}{\textbf{Total}} & \multicolumn{1}{l}{\textbf{0.69 / 0.27 / 1.41 / 0.25 (237)}} & \multicolumn{1}{l}{\textbf{0.05 / 0.02 / 0.10 / 0.02 (237)}} \\
    \hline
    \end{tabular}
    \caption{Travel speeds for children measured on the spiral staircases during the evacuation drills.}
    \label{tab:speed_stair_spiral}
\end{table}
\begin{table}[hbt!]
    \centering
    \footnotesize
    \begin{tabular}{R{1.3cm}|R{1.5cm}|R{5.3cm}|R{5.3cm}}
    \hline
      \textbf{Age group} & \textbf{Density interval [m$^2$\,m$^{-2}$]} & \textbf{Travel speed \newline (mean/min/max/SD) [m$\cdot$s$^{-1}$] (data points)} & \textbf{Density (mean/min/max/SD) [m$^2$\,m$^{-2}$] (data points)} \\
    \hline
    \multirow{2}{=}{Mixed} & 0.00--0.05 & 0.67 / 0.31 / 1.01 / 0.14 (48) &  0.04 / 0.02 / 0.05 / 0.01 (48)  \\
     \cline{2-4}
     & 0.06--0.10 & 0.58 / 0.31 / 0.99 / 0.19 (66) &  0.07 / 0.05 / 0.10 / 0.01 (66)  \\
    \hline
    \multicolumn{2}{l}{\textbf{Total}} & \multicolumn{1}{l}{\textbf{0.61 / 0.31 / 1.01 / 0.18 (114)}} & \multicolumn{1}{l}{\textbf{0.05 / 0.02 / 0.10 / 0.02 (114)}} \\
    \hline
    \end{tabular}
    \caption{Travel speeds for children measured on the spiral staircases on Travel path~1 (tread depth = 300~mm, slope of the path = 32.3$^{\circ}$).}
    \label{tab:speed_stair_spiral_1}
\end{table}
\begin{table}[hbt!]
    \centering
    \footnotesize
    \begin{tabular}{R{1.3cm}|R{1.5cm}|R{5.3cm}|R{5.3cm}}
    \hline
      \textbf{Age group} & \textbf{Density interval [m$^2$\,m$^{-2}$]} & \textbf{Travel speed \newline (mean/min/max/SD) [m$\cdot$s$^{-1}$] (data points)} & \textbf{Density (mean/min/max/SD) [m$^2$\,m$^{-2}$] (data points)} \\
    \hline
    \multirow{2}{=}{Mixed} & 0.00--0.05 & 0.89 / 0.36 / 1.41 / 0.27 (59) &  0.04 / 0.02 / 0.05 / 0.01 (59)  \\
     \cline{2-4}
     & 0.06--0.10 & 0.65 / 0.27 / 1.29 / 0.23 (64) &  0.07 / 0.05 / 0.10 / 0.01 (64)  \\
    \hline
    \multicolumn{2}{l}{\textbf{Total}} & \multicolumn{1}{l}{\textbf{0.76 / 0.27 / 1.41 / 0.27 (123)}} & \multicolumn{1}{l}{\textbf{0.05 / 0.02 / 0.10 / 0.02 (123)}} \\
    \hline
    \end{tabular}
    \caption{Travel speeds for children measured on the spiral staircases on Travel path~2 (tread depth = 450~mm, slope of the path = 22.3$^{\circ}$).}
    \label{tab:speed_stair_spiral_2}
\end{table}
\begin{table}[hbt!]
    \centering
    \footnotesize
    \begin{tabular}{R{1.3cm}|R{1.5cm}|R{5.3cm}|R{5.3cm}}
    \hline
      \textbf{Age group}  & {\textbf{Density interval [m$^2$\,m$^{-2}$]}} & \textbf{Specific flow \newline (mean/min/max/SD) [pers$\cdot$s$^{-1}$\,m$^{-1}$] (data points)} & \textbf{Density (mean/min/max/SD) [m$^2$\,m$^{-2}$] (data points)} \\
    \hline
    \multirow{2}{=}{Mixed}  & 0.00--0.05 & 1.78 / 1.21 / 3.64 / 0.77 (107) &  0.04 / 0.02 / 0.05 / 0.01 (107)  \\
     \cline{2-4}
       & 0.06--0.10 & 2.14 / 1.21 / 3.64 / 0.79 (130) & 0.07 / 0.05 / 0.10 / 0.01 (130)  \\
    \hline
    \multicolumn{2}{l}{\textbf{Total}} & \multicolumn{1}{l}{\textbf{1.98 / 1.21 / 3.64 / 0.77 (237)}} & \multicolumn{1}{l}{\textbf{0.05 / 0.02 / 0.10 / 0.02 (237)}} \\
    \hline
    \end{tabular}
    \caption{Specific flows for children measured on the spiral staircases during the evacuation drills.}
    \label{tab:flow_stair_spiral}
\end{table}

\clearpage\newpage
\section{Complementary material for the results on movement in doorways}
\label{app:door}
\setcounter{table}{0}
\renewcommand\thetable{\Alph{section}.\arabic{table}}
\begin{table}[hbt!]
    \centering
    \footnotesize
    \begin{tabular}{R{1.3cm}|R{1.5cm}|R{5.3cm}|R{5.3cm}}
    \hline
      \textbf{Age group} & \textbf{Density interval [m$^2$\,m$^{-2}$]}  & \textbf{Travel speed \newline (mean/min/max/SD) [m$\cdot$s$^{-1}$] (data points)} & \textbf{Density \newline (mean/min/max/SD) [m$^2$\,m$^{-2}$] (data points)} \\
    \hline
    \multirow{4}{=}{Junior} & 0.00--0.05 & 1.13 / 0.40 / 2.78 / 0.49 (48) & 0.03 / 0.03 / 0.04 / 0.00 (48)  \\
    \cline{2-4}
    & 0.06--0.10 & 0.88 / 0.12 / 2.00 / 0.34 (155) & 0.07 / 0.06 / 0.09 / 0.01 (155)  \\
    \cline{2-4}
    & 0.11--0.15 & 0.68 / 0.27 / 1.30 / 0.31 (68) & 0.11 / 0.11 / 0.14 / 0.01 (68)  \\ 
        \cline{2-4}
    & 0.16--0.20 & 1.01 / 0.76 / 1.20 / 0.23 (3) & 0.17 / 0.16 / 0.18 / 0.01 (3)  \\
    \hline
    \multirow{3}{=}{Senior} & 0.00--0.05 &  1.39 / 0.67 / 3.45 / 0.48 (54) &  0.03 / 0.03 / 0.04 / 0.00 (54)  \\
    \cline{2-4}
    & 0.06--0.10 & 1.06 / 0.33 / 2.13 / 0.35 (213) & 0.07 / 0.06 / 0.10 / 0.01 (213)  \\
    \cline{2-4}
    & 0.11--0.15 & 0.80 / 0.37 / 1.92 / 0.25 (79) & 0.12 / 0.11 / 0.14 / 0.01 (79)  \\    
    \hline
    \multirow{2}{=}{Senior+} & 0.06--0.10 & 0.64 / 0.51 / 0.78 / 0.10 (7)  & 0.09 / 0.09 / 0.09 / 0.00 (7) \\
    \cline{2-4}
    & 0.11--0.15 & 0.71 / 0.50 / 1.67 / 0.29 (15) & 0.12 / 0.11 / 0.15 / 0.01 (15)  \\        
    \hline
    \multirow{4}{=}{Mixed} & 0.00--0.05 & 1.73 / 0.49 / 3.70 / 0.73 (55) & 0.03 / 0.03 / 0.04 / 0.00 (55)  \\
    \cline{2-4}
    & 0.06--0.10 & 1.06 / 0.38 / 2.50 / 0.47 (95) & 0.07 / 0.06 / 0.10 / 0.01 (95)  \\
    \cline{2-4}
    & 0.11--0.15 & 0.67 / 0.27 / 1.25 / 0.19 (76) & 0.12 / 0.11 / 0.14 / 0.01 (76)  \\ 
        \cline{2-4}
    & 0.16--0.20 & 0.59 / 0.33 / 0.83 / 0.19 (6) & 0.16 / 0.16 / 0.16 / 0.00 (6)  \\
    \hline
    \multicolumn{2}{l}{\textbf{Total}}  & \multicolumn{1}{l}{\textbf{0.99 / 0.12 / 3.70 / 0.47 (874)}} & \multicolumn{1}{l}{\textbf{0.07 / 0.03 / 0.18 / 0.03 (874)}} \\
    \hline
    \end{tabular}
    \caption{Travel speeds for children measured in the doorways during the evacuation drills (the values in the row \quotes{Total} indicated the results over the total sample, i.e.,~all age groups and density intervals).}
    \label{tab:speed_door}
\end{table}
\begin{table}[hbt!]
    \centering
    \footnotesize
    \begin{tabular}{R{1.3cm}|R{1.5cm}|R{5.3cm}|R{5.3cm}}
    \hline
      \textbf{Age group} & \textbf{Density interval [m$^2$\,m$^{-2}$]}  & \textbf{Travel speed \newline (mean/min/max/SD) [m$\cdot$s$^{-1}$] (data points)} & \textbf{Density \newline (mean/min/max/SD) [m$^2$\,m$^{-2}$] (data points)} \\
    \hline
    \multirow{4}{=}{Junior} & 0.00--0.05 & 0.92 / 0.40 / 1.59 / 0.30 (35) & 0.03 / 0.03 / 0.04 / 0.00 (35)  \\
    \cline{2-4}
    & 0.06--0.10 & 0.84 / 0.12 / 2.00 / 0.31 (143) & 0.07 / 0.06 / 0.09 / 0.01 (143)  \\
    \cline{2-4}
    & 0.11--0.15 & 0.68 / 0.27 / 1.30 / 0.31 (68) & 0.11 / 0.11 / 0.14 / 0.01 (68)  \\ 
        \cline{2-4}
    & 0.16--0.20 & 1.01 / 0.76 / 1.20 / 0.23 (3) & 0.17 / 0.16 / 0.18 / 0.01 (3)  \\
    \hline
    \multirow{3}{=}{Senior} & 0.00--0.05 &  1.09 / 0.67 / 1.35 / 0.19 (25) &  0.03 / 0.03 / 0.04 / 0.00 (25)  \\
    \cline{2-4}
    & 0.06--0.10 & 0.96 / 0.33 / 2.00 / 0.26 (168) & 0.07 / 0.06 / 0.10 / 0.01 (168)  \\
    \cline{2-4}
    & 0.11--0.15 & 0.79 / 0.37 / 1.92 / 0.24 (78) & 0.12 / 0.11 / 0.14 / 0.01 (78)  \\    
    \hline
    \multirow{2}{=}{Senior+} & 0.06--0.10 & 0.64 / 0.51 / 0.78 / 0.10 (7)  & 0.09 / 0.09 / 0.09 / 0.00 (7) \\
    \cline{2-4}
    & 0.11--0.15 & 0.71 / 0.50 / 1.67 / 0.29 (15) & 0.12 / 0.11 / 0.15 / 0.01 (15)  \\        
    \hline
    \multirow{4}{=}{Mixed} & 0.00--0.05 & 0.95 / 0.49 / 1.32 / 0.24 (15) & 0.03 / 0.03 / 0.04 / 0.00 (15)  \\
    \cline{2-4}
    & 0.06--0.10 & 0.84 / 0.38 / 1.35 / 0.23 (71) & 0.07 / 0.06 / 0.09 / 0.01 (71)  \\
    \cline{2-4}
    & 0.11--0.15 & 0.67 / 0.27 / 1.25 / 0.19 (76) & 0.12 / 0.11 / 0.14 / 0.01 (76)  \\ 
        \cline{2-4}
    & 0.16--0.20 & 0.59 / 0.33 / 0.83 / 0.19 (6) & 0.16 / 0.16 / 0.16 / 0.00 (6)  \\
    \hline
    \multicolumn{2}{l}{\textbf{Total}}  & \multicolumn{1}{l}{\textbf{0.84 / 0.12 / 2.00 / 0.29 (710)}} & \multicolumn{1}{l}{\textbf{0.08 / 0.03 / 0.18 / 0.03 (710)}} \\
    \hline
    \end{tabular}
    \caption{Walking travel speeds for children measured in the doorways during the evacuation drills (the values in the row \quotes{Total} indicated the results over the total sample, i.e.,~all age groups and density intervals).}
    \label{tab:speed_door_walk}
\end{table}
\begin{table}[hbt!]
    \centering
    \footnotesize
    \begin{tabular}{R{1.3cm}|R{1.5cm}|R{5.3cm}|R{5.3cm}}
    \hline
      \textbf{Age group} & \textbf{Density interval [m$^2$\,m$^{-2}$]}  & \textbf{Travel speed \newline (mean/min/max/SD) [m$\cdot$s$^{-1}$] (data points)} & \textbf{Density \newline (mean/min/max/SD) [m$^2$\,m$^{-2}$] (data points)} \\
    \hline
    \multirow{2}{=}{Junior} & 0.00--0.05 & 1.70 / 1.03 / 2.78 / 0.48 (13) & 0.03 / 0.03 / 0.03 / 0.00 (13)  \\
    \cline{2-4}
    & 0.06--0.10 & 1.35 / 0.73 / 1.75 / 0.32 (17) & 0.06 / 0.06 / 0.08 / 0.01 (17)  \\
    \hline
    \multirow{3}{=}{Senior} & 0.00--0.05 &  1.64 / 0.83 / 3.45 / 0.52 (29) &  0.03 / 0.03 / 0.03 / 0.00 (29)  \\
    \cline{2-4}
    & 0.06--0.10 & 1.45 / 0.71 / 2.13 / 0.35 (45) & 0.06 / 0.06 / 0.08 / 0.01 (45)  \\
    \cline{2-4}
    & 0.11--0.15 & 0.79 / 0.37 / 1.92 / 0.24 (78) & 0.12 / 0.11 / 0.14 / 0.01 (78)  \\    
    \hline
    \multirow{2}{=}{Senior+} & 0.06--0.10 & 0.64 / 0.51 / 0.78 / 0.10 (7)  & 0.09 / 0.09 / 0.09 / 0.00 (7) \\
    \cline{2-4}
    & 0.11--0.15 & 1.43 / 1.43 / 1.43 / - (1) & 0.14 / 0.14 / 0.14 / - (1)  \\        
    \hline
    \multirow{2}{=}{Mixed} & 0.00--0.05 & 2.02 / 1.18 / 3.70 / 0.63 (40) & 0.03 / 0.03 / 0.03 / 0.00 (40)  \\
    \cline{2-4}
    & 0.06--0.10 & 1.70 / 1.03 / 2.50 / 0.42 (24) & 0.07 / 0.06 / 0.10 / 0.02 (24)  \\
    \hline
    \multicolumn{2}{l}{\textbf{Total}}  & \multicolumn{1}{l}{\textbf{1.67 / 0.71 / 3.70 / 0.52 (164)}} & \multicolumn{1}{l}{\textbf{0.04 / 0.02 / 0.14 / 0.02 (164)}} \\
    \hline
    \end{tabular}
    \caption{Running travel speeds for children measured in the doorways during the evacuation drills (the values in the row \quotes{Total} indicated the results over the total sample, i.e.,~all age groups and density intervals).}
    \label{tab:speed_door_run}
\end{table}
\begin{table}[hbt!]
    \centering
    \footnotesize
    \begin{tabular}{R{1.3cm}|R{1.5cm}|R{5.3cm}|R{5.3cm}}
    \hline
      \textbf{Age group}  & {\textbf{Density interval [m$^2$\,m$^{-2}$]}} & \textbf{Specific flow \newline (mean/min/max/SD) [pers$\cdot$s$^{-1}$\,m$^{-1}$] (data points)} & \textbf{Density (mean/min/max/SD) [m$^2$\,m$^{-2}$] (data points)} \\
    \hline
    \multirow{4}{=}{Junior}  & 0.00--0.05 & 1.32 / 0.71 / 3.75 / 0.42 (49) & 0.03 / 0.03 / 0.05 / 0.00 (49)  \\
     \cline{2-4}
                             & 0.06--0.10 & 2.55 / 1.00 / 6.25 / 0.88 (165) & 0.07 / 0.05 / 0.09 / 0.01 (165)  \\
     \cline{2-4}
                             & 0.11--0.15 & 2.89 / 1.11 / 6.25 / 1.21 (81) & 0.11 / 0.11 / 0.14 / 0.01 (81)  \\    
     \cline{2-4}
                             & 0.16--0.20 & 2.27 / 2.22 / 2.35 / 0.07 (5) & 0.16 / 0.15 / 0.18 / 0.01 (5)  \\  
    \hline
    \multirow{4}{=}{Senior}  & 0.00--0.05 & 1.53 / 1.11 / 3.75 / 0.62 (54) & 0.03 / 0.03 / 0.04 / 0.00 (54)  \\
     \cline{2-4}
                             & 0.06--0.10 & 2.66 / 1.11 / 5.56 / 0.79 (228) & 0.07 / 0.06 / 0.10 / 0.01 (228)  \\
     \cline{2-4}
                             & 0.11--0.15 & 3.23 / 1.11 / 5.71 / 1.03 (82) & 0.12 / 0.11 / 0.14 / 0.01 (82)  \\    
     \cline{2-4}
                             & 0.16--0.20 & 3.33 / 3.33 / 3.33 / 0.00 (2) & 0.16 / 0.16 / 0.16 / 0.00 (2)  \\ 
    \hline
    \multirow{2}{=}{Senior+} & 0.06--0.10 & 2.70 / 1.11 / 3.33 / 0.87 (7) & 0.09 / 0.09 / 0.09 / 0.00 (7)  \\
     \cline{2-4}
                             & 0.11--0.15 & 3.26 / 2.22 / 3.33 / 0.29 (15) & 0.12 / 0.11 / 0.14 / 0.01 (22)  \\ 
    \hline
    \multirow{4}{=}{Mixed}  & 0.00--0.05 & 1.47 / 0.11 / 5.00 / 0.80 (57) & 0.03 / 0.03 / 0.04 / 0.00 (57)  \\
     \cline{2-4}
                             & 0.06--0.10 & 2.29 / 0.91 / 5.00 / 0.89 (109) & 0.07 / 0.05 / 0.10 / 0.01 (109)  \\
     \cline{2-4}
                             & 0.11--0.15 & 2.99 / 1.11 / 6.25 / 0.15 (90) & 0.12 / 0.11 / 0.15 / 0.01 (90)  \\    
     \cline{2-4}
                             & 0.16--0.20 & 3.40 / 2.22 / 6.25 / 1.05 (11) & 0.16 / 0.16 / 0.18 / 0.01 (11)  \\ 
    \hline
    \multicolumn{2}{l}{\textbf{Total}} & \multicolumn{1}{l}{\textbf{2.51 / 0.11 / 6.25 / 1.04 (955)}} & \multicolumn{1}{l}{\textbf{0.08 / 0.03 / 0.18 / 0.03 (955)}} \\
    \hline
    \end{tabular}
    \caption{Specific flows for children measured in the doorways during the evacuation drills (the values in the row \quotes{Total} indicated the results over the total sample, i.e.,~all age groups and density intervals).}
    \label{tab:flow_door}
\end{table}
\clearpage\newpage
\section{Complementary material for the results on movement behaviour}
\label{app:beh}
\setcounter{table}{0}
\renewcommand\thetable{\Alph{section}.\arabic{table}}
\begin{table}[hbt!]
    \centering
    \footnotesize
    \begin{tabular}{R{3.7cm}|R{2.0cm}|R{1.7cm}|R{1.7cm}|R{1.7cm}|R{1.7cm}}
    \hline
    \multirow{2}{=}{\textbf{Physical assistance provided}} & \multicolumn{5}{R{7,5cm}}{\textbf{Frequency [\%] (data points)}} \\
    \cline{2-6}
    & \textbf{Total} & \textbf{Junior} & \textbf{Senior} & \textbf{Senior+} & \textbf{Mixed} \\
    \hline
    \multicolumn{6}{l}{\textbf{Corridors}} \\
    \hline
    No physical assistance & 88.1 (1099) & 88.5 (314) & 90.0 (448) & 100.0 (132) & 77.9 (205) \\
    \hline
    Gentle pushing & 6.2 (77) & 4.5 (16) & 9.4 (47) & 0.0 (0) & 5.3 (14) \\
    \hline
    Physical contact needed & 1.3 (16) & 2.5 (9) & 0.0 (0) & 0.0 (0) & 2.7 (7)  \\
    \hline
    Hand holding & 4.2 (53) & 4.5 (16) & 0.6 (3) & 0.0 (0) & 12.9 (34) \\
    \hline
    Carried & 0.2 (3) & 0.0 (0) & 0.0 (0) & 0.0 (0) & 1.1 (3) \\
\hline
    \multicolumn{6}{l}{\textbf{Straight staircases}} \\
    \hline
    No physical assistance & 92.2 (1414) & 90.4 (310) & 95.1 (796) & 94.5 (104) & 84.0 (204) \\
    \hline
    Gentle pushing & 1.6 (24) & 0.0 (0) & 0.8 (7) & 0.0 (0) & 7.0 (17) \\
    \hline
    Physical contact needed & 0.5 (8) & 1.2 (4) & 0.0 (0) & 0.0 (0) & 1.6 (4)  \\
    \hline
    Hand holding & 5.7 (87) & 8.5 (29) & 4.1 (34) & 5.5 (6) & 7.4 (18) \\
    \hline
    Carried & 0.0 (0) & 0.0 (0) & 0.0 (0) & 0.0 (0) & 0.0 (0) \\
\hline
    \multicolumn{6}{l}{\textbf{Spiral staircases}} \\
    \hline
    No physical assistance & 97.6 (321) & N/A & N/A & N/A & 97.6 (321) \\
    \hline
    Gentle pushing & 0.0 (0) & N/A & N/A & N/A & 0.0 (0) \\
    \hline
    Physical contact needed & 0.9 (3) & N/A & N/A & N/A & 0.9 (3)  \\
    \hline
    Hand holding & 1.5 (5) & N/A & N/A & N/A & 1.5 (5) \\  
    \hline
    Carried & 0.0 (0) & 0.0 (0) & 0.0 (0) & 0.0 (0) & 0.0 (0) \\
\hline
    \multicolumn{6}{l}{\textbf{Doorways}} \\
    \hline
    No physical assistance & 86.6 (806) & 86.2 (238) & 88.3 (323) & 90.9 (20) & 84.3 (225) \\
    \hline
    Gentle pushing & 8.1 (75) & 6.2 (17) & 8.5 (31) & 9.1 (2) & 9.4 (25) \\
    \hline
    Physical contact needed & 2.3 (21) & 2.9 (8) & 0.5 (2) & 0.0 (0) & 4.1 (11)  \\
    \hline
    Hand holding & 3.1 (29) & 4.7 (13) & 2.7 (10) & 0.0 (0) & 2.2 (6) \\
    \hline
    Carried & 0.0 (0) & 0.0 (0) & 0.0 (0) & 0.0 (0) & 0.0 (0) \\
    \hline
    \end{tabular}
    \caption{Levels of physical assistance provided to children in different age groups on different parts of evacuation routes during the evacuation drills.}
    \label{tab:mov_contact}
\end{table}
\begin{table}[hbt!]
    \centering
    \footnotesize
    \begin{tabular}{R{3.8cm}|R{2cm}|R{1.7cm}|R{1.7cm}|R{1.7cm}|R{1.6cm}}
    \hline
    \multirow{2}{=}{\textbf{Hand holding}} & \multicolumn{5}{R{7.5cm}}{\textbf{Frequency [\%] (data points)}} \\
    \cline{2-6}
    & \textbf{Total} & \textbf{Junior} & \textbf{Senior} & \textbf{Senior+} & \textbf{Mixed} \\
    \hline
    \multicolumn{6}{l}{\textbf{Corridors}} \\
    \hline
    No hand holding & 49.4 (617) & 36.9 (131) & 32.7 (163) & 78.8 (104) & 83.3 (219) \\
    \hline
    With another child & 46.3 (578) & 58.6 (208) & 66.7 (332) & 21.2 (28) & 3.8 (10) \\
    \hline
    With a staff member & 3.8 (48) & 3.4 (12) & 0.4 (2) & 0.0 (0) & 12.9 (34)  \\
    \hline
    With another child and a staff member & 0.4 (5) & 1.1 (4) & 0.2 (1) & 0.0 (0) & 0.0 (0) \\
\hline
    \multicolumn{6}{l}{\textbf{Straight staircases}} \\
    \hline
   No hand holding & 63.0 (966) & 64.1 (220) & 50.8 (425) & 87.3 (96) & 92.6 (225) \\
    \hline
    With another child & 31.3 (480) & 27.4 (94) & 45.2 (378) & 7.3 (8) & 0.0 (0) \\
    \hline
    With a staff member & 5.2 (79) & 6.7 (23) & 3.8 (32) & 5.5 (6) & 7.4 (18)  \\
    \hline
    With another child and a staff member & 0.5 (8) & 1.7 (6) & 0.2 (2) & 0.0 (0) & 0.0 (0) \\
\hline
    \multicolumn{6}{l}{\textbf{Spiral staircases}} \\
    \hline
    No hand holding & 78.1 (257) & N/A & N/A & N/A & 78.1 (257) \\
    \hline
    With another child & 20.4 (67) & N/A & N/A & N/A & 20.4 (67) \\
    \hline
    With a staff member & 1.5 (5) & N/A & N/A & N/A & 1.5 (5)  \\
    \hline
    With another child and a staff & 0.0 (0) & N/A & N/A & N/A & 0.0 (0) \\
\hline
    \multicolumn{6}{l}{\textbf{Doorways}} \\
    \hline
    No hand holding & 59.3 (552) & 49.6 (137) & 41.5 (152) & 63.3 (14) & 93.3 (249) \\
    \hline
    With another child & 37.6 (350) & 45.7 (126) & 55.7 (204) & 36.4 (8) & 4.5 (12) \\
    \hline
    With a staff member & 2.6 (24) & 2.9 (8) & 2.7 (10) & 0.0 (0) & 2.2 (6)  \\
    \hline
    With another child and a staff member & 0.5 (5) & 1.8 (5) & 0.0 (0) & 0.0 (0) & 0.0 (0) \\
\hline
  \end{tabular}
\caption{Hand holding observed in different age groups and in different parts of the evacuation routes during evacuation drills.}
\label{tab:mov_hand}
\end{table}
\begin{table}[hbt!]
    \centering
    \footnotesize
    \begin{tabular}{R{3.9cm}|R{1.8cm}|R{1.7cm}|R{1.7cm}|R{1.7cm}|R{1.7cm}}
    \hline
    \multirow{2}{=}{\textbf{Use of handrails on straight staircases}} & \multicolumn{5}{R{7.5cm}}{\textbf{Frequency [\%] (data points)}} \\
    \cline{2-6}
    & \textbf{Total} & \textbf{Junior} & \textbf{Senior} & \textbf{Senior+} & \textbf{Mixed} \\
\hline
    \multicolumn{6}{l}{\textbf{Only standard handrail available}} \\
    \hline
    Without the use of handrails & 22.7 (83) & 53.9 (41) & 45.3 (34) & 2.7 (3) & 4.8 (5) \\
    \hline
    Standard handrail & 76.5 (280) & 42.1 (32) & 54.7 (41) & 97.3 (107) & 95.2 (100) \\
    \hline
    Baluster or other support & 0.8 (3) & 3.9 (3) & 0.0 (0) & 0.0 (0) & 0.0 (0) \\    
\hline
    \multicolumn{6}{l}{\textbf{Only children's handrail available}} \\
    \hline
    Without the use of handrails & 73.9 (34) & N/A & N/A & N/A & 73.9 (34) \\
    \hline
    Children's handrail & 26.1 (12) & N/A & N/A & N/A & 26.1 (12) \\
    \hline
    Baluster or other support & 0.0 (0) & N/A & N/A & N/A & 0.0 (0) \\
\hline
    \multicolumn{6}{l}{\textbf{Both standard and children's handrail available}} \\
    \hline
    Without the use of handrails & 34.1 (346) & 18.6 (37) & 35.8 (259) & N/A & 54.3 (50) \\
    \hline
    Standard handrail & 33.1 (336) & 57.3 (114) & 30.1 (218) & N/A & 4.3 (4) \\
    \hline    
    Children's handrail & 29.8 (302) & 20.1 (40) & 30.9 (224) & N/A & 41.3 (38) \\
    \hline
    Baluster or other support & 3.1 (31) & 4.0 (8) & 3.2 (23) & N/A & 0.0 (0) \\
\hline
  \end{tabular}
   \caption{Handrail use observed for different age groups on straight staircases with respect to different types of handrails available}
    \label{tab:mov_handrails}
\end{table}

\biboptions{sort&compress}
\end{document}